\documentclass[journal]{IEEEtran}
\usepackage{cite}
\usepackage{amsmath,amssymb,amsfonts}
\usepackage{graphicx}
\usepackage{textcomp}
\usepackage{caption}
\usepackage{xspace}
\usepackage{siunitx}
\usepackage{subfigure}
\usepackage{ulem}
\usepackage{url}
\usepackage{hyperref,xcolor}
\usepackage{algpseudocode}
\usepackage{algorithm}
\usepackage{algorithmicx}
\hypersetup{colorlinks=true, linkcolor=blue!80!black, urlcolor=blue!80!black, citecolor=blue!80!black}

\usepackage[nameinlink]{cleveref}
\crefname{figure}{Fig.}{Figs.}
\Crefname{figure}{Fig.}{Figs.}

\newif\iffinal

\iffinal
    \newcommand\joaquin[1]{}
    
    \newcommand{\red}[1]{}
\else
    \newcommand\joaquin[1]{{\color{blue}[Joaquin: #1]}}
    
    \newcommand{\red}[1]{\textcolor{red}{#1}}
\fi

\newcommand{\Erppp}{Er\textsuperscript{3+}}

\def\BibTeX{{\rm B\kern-.05em{\sc i\kern-.025em b}\kern-.08em
    T\kern-.1667em\lower.7ex\hbox{E}\kern-.125emX}}
    
\begin{document}
\bstctlcite{IEEEexample:BSTcontrol}

\title{InterQnet: A Heterogeneous Full-Stack Approach to Co-designing Scalable Quantum Networks}

\author{
    \IEEEauthorblockN{
        Joaquin Chung$^{1}$,
        Daniel Dilley$^{2}$,
        Ely Eastman$^{3}$,
        Alvin Gonzales$^{2}$,
        Kara Hokenstad$^{3}$,
        Md Shariful Islam$^{1}$,
        Varun Jorapur$^{4}$,
        Joseph Petrullo$^{5}$,
        Andy C.~Y.~Li$^{6}$,
        Bikun Li$^{5}$,
        Vasileios Niaouris$^{7}$,
        Anirudh Ramesh$^{3}$,
        Ansh Singal$^{3}$,
        Zeyu Ye$^{4,5}$,
        Caitao Zhan$^{1}$,
        Michael Bishof$^{4}$,
        Eric Chitambar$^{8}$,
        Jacob P. Covey$^{9}$,
        Alan Dibos$^{7}$,
        Xu Han$^{10}$,
        Liang Jiang$^{5}$,
        Prem Kumar$^{3}$,
        Jeffrey Larson$^{2}$,
        Zain H. Saleem$^{2}$, and
        Rajkumar Kettimuthu$^{1}$
    }
    
    \IEEEauthorblockA{
        $^{1}$ Data Science and Learning Division, Argonne National Laboratory, Lemont, IL 60439 USA\\
        $^{2}$ Mathematics and Computer Science Division, Argonne National Laboratory, Lemont, IL 60439 USA\\
        $^{3}$ Northwestern University, Evanston, IL 60208 USA\\
        $^{4}$ Physics Division, Argonne National Laboratory, Lemont, IL 60439 USA\\
        $^{5}$ The University of Chicago, Chicago, IL 60637 USA\\
        $^{6}$ Fermi National Accelerator Laboratory, Batavia, IL 60510 USA\\
        $^{7}$ Q-NEXT, Argonne National Laboratory, Lemont, IL 60439 USA\\
        $^{8}$ Electrical \& Computer Eng. Department, University of Illinois Urbana-Champaign, Urbana, IL 61801 USA\\
        $^{9}$ Physics Department, University of Illinois Urbana-Champaign, Urbana, IL 61801 USA\\
        $^{10}$ Center for Nanoscale Materials, Argonne National Laboratory, Lemont, IL 60439 USA\\
    }

    \thanks{This material is based upon work supported by the U.S.~Department of Energy, Office Science, Advanced Scientific Computing Research (ASCR) program under contract number DE-AC02-06CH11357 as part of the InterQnet quantum networking project. This manuscript has been authored by Fermi Forward Discovery Group, LLC under Contract No. 89243024CSC000002 with the U.S. Department of Energy, Office of Science, Office of High Energy Physics.}
}



\maketitle

\begin{abstract}
Quantum communications have progressed significantly, moving from a theoretical concept to small-scale experiments to recent metropolitan-scale demonstrations.
As the technology matures, it is expected to revolutionize quantum computing in much the same way that classical networks revolutionized classical computing.
Quantum communications will also enable breakthroughs in quantum sensing, metrology, and other areas.
However, scalability has emerged as a major challenge, particularly in terms of the number and heterogeneity of nodes, the distances between nodes, the diversity of applications, and the scale of user demand.
This paper describes InterQnet, a multidisciplinary project that advances scalable quantum communications through a comprehensive approach that improves devices, error handling, and network architecture. 
InterQnet has a two-pronged strategy to address scalability challenges: 
\textit{InterQnet-Achieve} focuses on practical realizations of heterogeneous quantum networks by building and then integrating first-generation quantum repeaters with error mitigation schemes and centralized automated network control systems. 
The resulting system will enable quantum communications between two heterogeneous quantum platforms through a third type of platform operating as a repeater node. 
\textit{InterQnet-Scale} focuses on a systems study of architectural choices for scalable quantum networks by developing forward-looking models of quantum network devices, advanced error correction schemes, and entanglement protocols.
Here we report our current progress toward achieving our scalability goals.
\end{abstract}



\section{Introduction} \label{sec:introduction}
Quantum networks hold enormous potential for ground-breaking advances in many areas of science and technology. 
Distributed quantum computing~\cite{CALEFFI2024110672} and sensing~\cite{zhang2021distributed} are among the important applications envisioned for quantum networks. 
As with the classical Internet, further applications not yet imagined can be expected to eclipse those currently envisioned, once scalable quantum networks become a reality. 

Yet despite the demonstration of entanglement distribution and teleportation of quantum states in laboratory-, campus-, and metropolitan-scale networks~\cite{LIU2025100551}, scaling quantum networks to longer distances and to multiple users and applications remains a major challenge.
The InterQnet project is advancing the state of the art in scalable quantum networks through work aimed at two interrelated overarching goals: (1) practical realization of a full-stack heterogeneous quantum network (\textbf{InterQnet-Achieve}) and (2) systems study of scalable quantum network architectures (\textbf{InterQnet-Scale}). By accomplishing these goals, \textit{InterQnet will provide fundamental building blocks and foundational architectural choices for national-scale quantum networks}.

\begin{figure}
    \centering
    \includegraphics[width=\columnwidth]{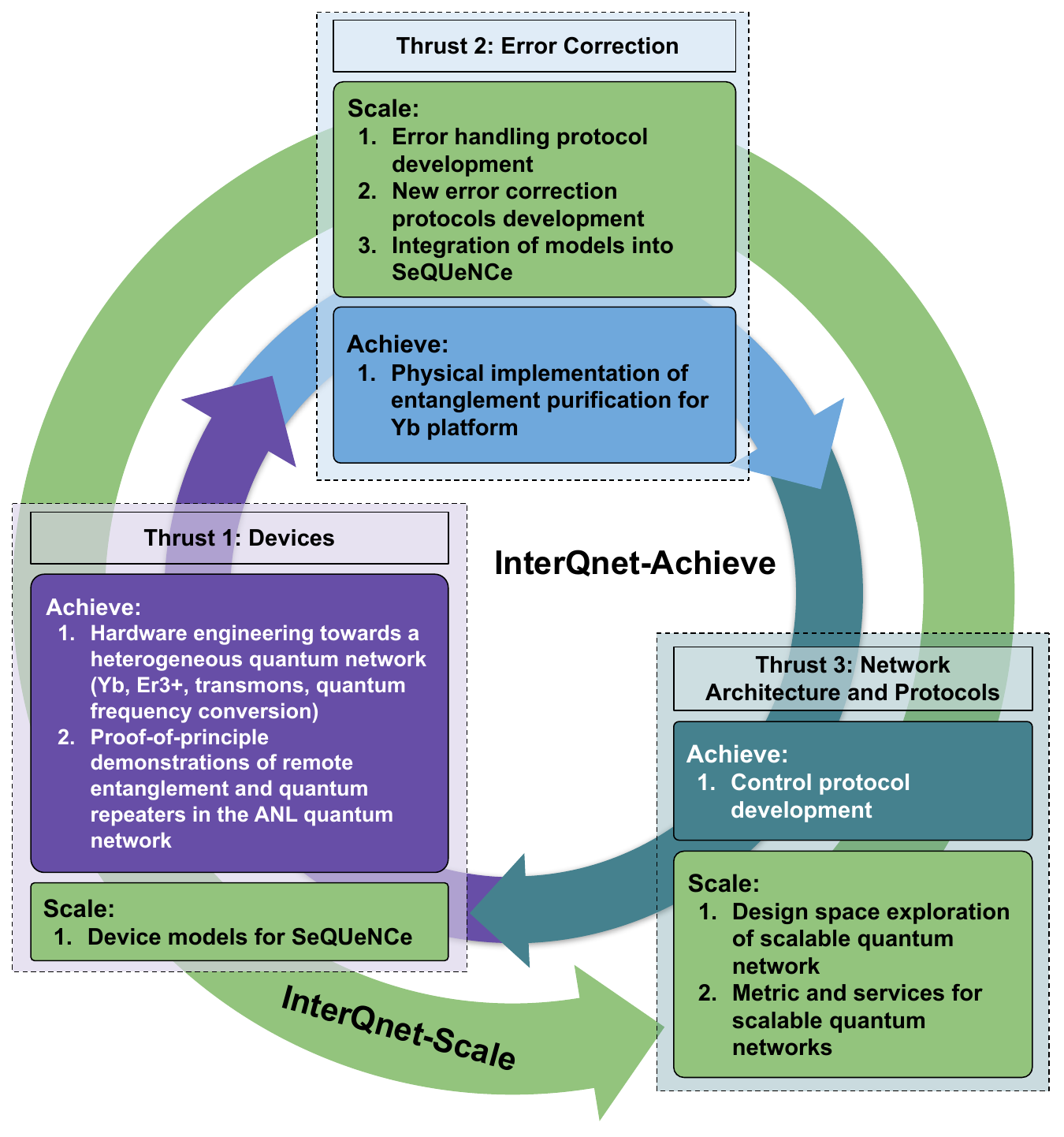}
    \caption{Organization of the InterQnet project, highlighting our systems approach to scalable quantum networks. The approach integrates devices, error correction methods, protocols, architectures, and simulation/experimentation into two co-design/integration cycles.}
    \label{fig:concept}
\end{figure}

In \textbf{InterQnet-Achieve} (inner circle in \Cref{fig:concept}), we are developing a repeater-based heterogeneous quantum network by co-designing (1) a set of devices including a prototype first-generation quantum repeater, microwave-optical quantum transducers, and optical quantum frequency converters; (2) related error mitigation strategies for communications; and (3) the necessary control and management protocols and interfaces, in a continuous cycle of development, validation, and optimization using physical testbeds and simulation studies.

While we use a centralized control and customized network protocols and error mitigation methods in \textbf{InterQnet-Achieve}, we investigate alternative quantum network architectures, sophisticated protocols, and advanced error correction methods in \textbf{InterQnet-Scale} (outer circle in \Cref{fig:concept}) 
by (1) implementing them in SeQUeNCe~\cite{sequence}---a scalable quantum network simulator;
(2) performing at-scale simulations of them by building models of future devices (in addition to the models for current devices in \textbf{InterQnet-Achieve}); 
and (3) utilizing the experimental testbeds to validate certain aspects of the simulations.

The remainder of the paper is organized as follows.
\Cref{sec:background} reviews the most recent experimental demonstrations of quantum networks and proposals of architectures for quantum networks, highlighting the main challenges for scalability.
\Cref{sec:approach} describes our approach to address scalability challenges.
As illustrated in \Cref{fig:concept}, InterQnet involves three thrusts: Heterogeneous Devices (\Cref{sec:devices}), Error Management (\Cref{sec:qec}), and Network Architecture and Protocols (\Cref{sec:net-arch}).
These sections, along with \Cref{sec:integration}, provide details on our progress toward the InterQnet goals.
\Cref{sec:conclusion} concludes the manuscript with an outlook for future achievements.
\section{Background} \label{sec:background}

Researchers have demonstrated early prototypes of campus- and metropolitan-scale quantum communications~\cite{valivarthi2016quantum,alshowkan2021,Pompili2022,thomas2024quantum,stolk2024entanglement,knaut2024entanglement,liu2024creation,craddock_automated_2024}.
However, unavoidable signal losses make quantum repeaters essential for long-range ($\gtrsim$\,100-kilometer) communication, with repeater-assisted entanglement distribution envisioned through quantum memories~\cite{lei2023quantum}.
Meanwhile, on the theory front, several quantum network architectures and protocol stacks have been proposed for a future quantum internet~\cite{kimble2008quantum}. 
For instance, Van Meter et al.~\cite{van-meter-2009} proposed a layered stack approach and later developed this into a recursive architecture~\cite{van2022quantum}. 
Wehner's group at TU Delft has also proposed layered models for quantum networks~\cite{dahlberg2019,kozlowski-2019} that roughly map the classical TCP/IP stack. 
While Van Meter's and Wehner's work focused on the distribution of bipartite entangled states on demand, Pirker and D\"ur~\cite{Pirker-2019} envisioned an architecture in which multipartite entangled states are distributed in advance and consumed in response to user requests.
However, early architecture proposals for a quantum internet impose stringent performance requirements on the underlying hardware devices.
Thus, a significant gap persists between the theory and practice of quantum networking. 
The following subsections provide details on efforts that attempt to close this gap by building a quantum networking testbed, defining novel quantum networking architectures, and evaluating architectural choices via simulations.

\subsection{Quantum Networking Testbeds}
Recent efforts to build and characterize deployed-fiber quantum network testbeds are making progress in bridging the gap between theory and practical implementation.
Pompili et al.\ demonstrated a phase-stabilized quantum network architecture based on three entangled nodes~\cite{pompili2021realization} and achieved entanglement delivery~\cite{Pompili2022} using the protocol stack proposed in~\cite{dahlberg2019}. 
Similarly, Stolk et al.~\cite{stolk2024entanglement} demonstrated heralded entanglement generation of predefined states between two diamond spin nodes in Delft and The Hague. The two nodes were connected to a midpoint station over a total distance of 25~km of deployed fiber, and photons were converted to the telecom L-band.
Knaut et al.~\cite{knaut2024entanglement} demonstrated nuclear spin entanglement distribution between two SiV nodes connected through 35~km of deployed fiber routed across the greater Boston metropolitan area, with quantum frequency conversion to the telecom O-band.
Similarly, Liu et al.~\cite{liu2024creation} demonstrated the creation of entanglement between two rubidium quantum memories separated by 12.5~km of deployed fiber in the city of Hefei.
Craddock et al.~\cite{craddock_automated_2024} used the GothamQ testbed in New York to demonstrate automated distribution of polarization-entangled photons over 34~km of underground deployed fiber.
The authors reported continuous distribution of entangled pairs for 15~days, with a network uptime of 99.84\%. 
The BARQNET testbed~\cite{bersin2024development} has provided valuable insights into characterizing and compensating for channel impairments including polarization and phase noise over a 50~km fiber loop, highlighting effective integration strategies for quantum memory compatibility.
The QUANT-NET testbed~\cite{monga2023quant,schon2024quant} is building a multinode quantum network connecting Lawrence Berkeley National Laboratory and UC Berkeley over deployed fiber, to implement entanglement swapping and teleportation of controlled-NOT gates between distant trapped-ion processors. 
In the interim, QUANT-NET has deployed essential infrastructure components such as Hong--Ou--Mandel interferometry and polarization drift compensation, controlled by a two-level control framework~\cite{quantunet-qce25}. 
 
\subsection{Quantum Network Architectures}
Early attempts to define an architecture and protocol stack for a quantum internet have tried to replicate the success of the classical Internet; thus, their proponents defined stacks~\cite{thomasreport,dahlberg2019,alshowkan2021} that map directly to the TCP/IP~\cite{forouzan2002tcp} stack.
As mentioned before, these architectures impose stringent requirements on hardware devices. These requirements include  asynchronous operations enabled by quantum memories with long coherence times and gate operations and measurements with high fidelity~\cite{van2022quantum} so protocols can run in a distributed fashion.
Taking a step toward TCP/IP-inspired architectures, Delle~Donne et al.~\cite{delle2025operating} presented QNodeOS, a platform-independent operating system for end nodes of a quantum network. QNodeOS enables multitasking and high-level application execution on heterogeneous quantum hardware.
However, current near-term quantum networks---of campus and metropolitan scale---will require centralized control and architectures that follow software-defined networking principles~\cite{martin2019sdn-qkd,ieqnet-arch2022,quantum-wrappers}.
Moreover, the current consensus is that these networks will serve entanglement requests as they arrive at the central controller. The QUANT-NET~\cite{quantunet-qce25} and Quantum Internet Alliance~\cite{beauchamp2025modularquantumnetworkarchitecture} groups have demonstrated implementations of this approach, and we adopt a centralized control architecture for our InterQnet-Achieve demonstration.

While on-demand entanglement distribution may become the architecture for long-distance quantum networks, architectures that continuously generate link-level entanglement deserve attention. 
In the context of distributed quantum computing, Talsma et al.~\cite{talsma2024continuously} analyzed a protocol for continuous entanglement distribution among nodes arranged in regular patterns (e.g., linear, honeycomb, square grid, and triangle lattice topologies). The study found that the number of swap operations required to increase the virtual neighborhood of a node (i.e., the number of entangled neighbors of a given node) increases as the coherence time of quantum memories decreases.
Gu et al.~\cite{gu2023esdi} proposed a framework called ESDI for long-distance entanglement scheduling and distribution in a general quantum network. By leveraging discrete-time simulations, the authors showed that ESDI can significantly reduce the average completion time of long-distance entanglement requests. 
Vardoyan et al.~\cite{vardoyan2021capacity} studied the capacity region of a switch that can generate bipartite or tripartite entanglement between end users connected to the switch in a star topology. The study showed that a set of randomized switching policies can outperform a time-division multiplexing  policy, even when quantum memory decoherence is considered.

Departing from the state of the art, Pirker et al.~\cite{pirker2025resource} proposed a shift from layered architectures to a resource-centric, task-based model. 
In this approach, applications define high-level objectives that are fulfilled through distributed workflows---called ``sagas''---that operate directly on quantum channels, classical messaging, and shared entanglement. 
This flexible model enables decentralized orchestration and emphasizes dynamic control, raising new questions about metrics, automation, and abstractions for scalable quantum networking.
In a more radical shift, Caleffi and Cacciapuoti have proposed a novel hierarchical quantum internet architecture centered around the concept of an entanglement-defined controller~\cite{caleffi2025quantuminternetarchitectureunlocking}.
Such a controller would enable scalable and efficient management of in-network operations such as the distribution and manipulation of quantum entanglement. 

\subsection{Simulations of Quantum Networks}
Simulators have been used to study architectural design choices of quantum networks, with repeater chains being a popular subject of study in the theory community.
For instance, Avis et al.~\cite{avis2023requirements} studied the hardware requirements of a quantum repeater node deployed on a hypothetical metropolitan-scale fiber network. The network connects the cities of Delft and Eindhoven (separated by 226.5~km), with the repeater located in the city of Utrecht. Besides realistic parameters of the fiber network, the authors considered the parameters for repeaters based on color centers in diamonds and trapped ion devices. Their results confirm the discrepancies between abstract/simplified and more realistic models. Moreover, the simulation results
provide minimal hardware requirements to build quantum repeaters based on the chosen hardware platforms.
Similarly, Jain et al.~\cite{jain2025trapped-ion} created accurate models of trapped ion quantum repeaters and simulated the performance of repeater chains using current and future parameters of the devices. The authors considered a hypothetical deployment over the ESnet network---a research and education network in the United States---and concluded that current trapped-ion platforms are unable to yield any usable entanglement fidelities and rates. Thus, device improvements, purification schemes, and quantum error correction mechanisms are needed for these platforms to become practical.

Other architectural design choices have been evaluated via simulations.
For example, Chan et al.~\cite{chan2025all-photonic} studied chains of all-photonic quantum repeaters.
Soon et al.~\cite{soon2024heterogeneous-links} investigated quantum networks composed of heterogeneous links (i.e., some links used the meet-in-the-middle protocol while others used the midpoint source protocol). Their simulations showed that while performance is dependent on link configuration, no significant degradation was observed when compared with homogeneous networks. 
 Koyama et al.~\cite{koyama2024switch} investigated the switched interconnects required to build modular quantum computers. The authors proposed designs for reducing loss and crosstalk while raising entanglement rates and fidelity.
Van Dam~\cite{vanDam2024bqc} et al. investigated the hardware requirements for executing verifiable blind quantum computation (BQC) using a trapped ion server and a distant measurement-only client, separated by a 50~km link.

\subsection{Open Scalability Challenges} \label{sec:open-challenges}
\begin{figure}
    \centering
    \includegraphics[width=0.7\columnwidth]{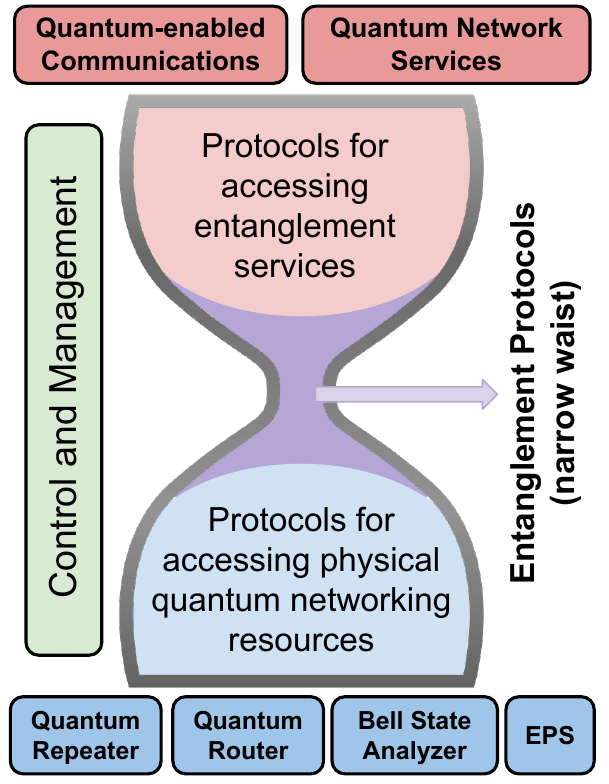}
    \caption{An entanglement-centric, hourglass-like view of quantum networking. This abstraction is provided for illustrative purposes and will not be evaluated in this work.} 
    \label{fig:qn-stack-hourglass}
\end{figure}
The transition from small-scale, isolated experiments to national-scale, heterogeneous quantum networks is hindered by limitations in scalability, control, and architectural understanding. 

\noindent\textbf{C1. Integration of heterogeneous technologies:}
Different qubit platforms excel at specific tasks, while having deficiencies in others.
Thus, integrating complementary qubit platforms into heterogeneous systems is a promising path toward scalability.
Furthermore, the ability to convert between different degrees of freedom in encoded photons~\cite{lu2025fully} and leverage diverse physical media (e.g., fiber optics and free space)~\cite{sheridan2025telecommunicationsfiberopticfreespacequantum} will greatly improve heterogeneous quantum networks.
Other groups besides InterQnet are also focusing on integrating heterogeneous nodes into their network testbeds. 
For instance, the QUANT-NET team is planning to demonstrate quantum state teleportation from a G-center node to a trapped-ion node~\cite{schon2024quant}. 
Similarly, the Quantum Internet Alliance is planning to interconnect trapped-ion end-nodes through a chain of rare-earth quantum repeaters~\cite{cussenot2025uniting}.
As seen in the literature review, quantum frequency conversion (QFC) is an important component for interconnecting matter-based  qubit platforms over telecom fiber~\cite{stolk2024entanglement,knaut2024entanglement}.
QFC is already challenging, but the difficulty increases when interconnecting heterogeneous platforms because one now must also consider photon waveform matching~\cite{cussenot2025uniting}.
In summary, \textit{interconnecting heterogeneous platforms over quantum networks with rates and fidelity sufficiently high for practical applications remains an open challenge.}

\noindent\textbf{C2. Control Frameworks for Quantum Networks:}
While long-term visions of a quantum internet propose architectures that resemble the TCP/IP stack and leverage distributed control~\cite{van2022quantum,kozlowski_rfc_2023}, recent testbed demonstrations focus on centralized control architectures~\cite{quantunet-qce25,beauchamp2025modularquantumnetworkarchitecture}.
Based on their experience implementing and evaluating a modular control architecture for quantum networks, Beauchamp et al.~\cite{beauchamp2025whitepaperquantuminternet} presented a collection of open questions and challenges for developing future quantum network control architectures and proposed potential solutions. Worth highlighting are the need for good admission control for user requests, computationally fast scheduling methods, and the need to know when to establish service agreements between the network and users.
Hayek et al.~\cite{hayek2025reviewsoftwaredesigningoperating} presented a review of software for designing and operating quantum networks, with a special focus on the network functions of the control plane. The authors identified areas that need further attention, such as topology management functions that can handle dynamic reconfiguration, standard interfaces for interoperability, and a management plane for resilience of the network.
In summary, \textit{advancing the control frameworks of quantum networks from automation of experiments to platforms that actually provide services to users remains an open challenge.}

\noindent\textbf{C3. Refined Architectural Understanding:}
Li et al.~\cite{li2024qiprotocolsurvey} presented a comprehensive survey on protocols for a quantum internet, from the perspective of layered models. The authors identified several challenges, such as increasing the realism of models for physical devices, establishing clear boundaries between layers---especially between physical and link, as well as between network and transport, and integrating quantum networks with existing infrastructure. 
The survey further emphasized the lack of consensus among layered model proposals. We argue that a useful lesson from the classical Internet is not the TCP/IP stack itself but its hourglass shape. A single narrow waist connects many protocols above and below. In quantum networks, entanglement protocols can play this role, accommodating heterogeneous physical platforms underneath and diverse applications on top (see \Cref{fig:qn-stack-hourglass}). 
\textit{Achieving such an hourglass-shaped stack will require clear abstractions for components; scalable error management; and a transversal layer for management, monitoring, and validation.}

These open challenges motivate the dual goals of InterQnet-Achieve and InterQnet-Scale described in \Cref{sec:approach}. Sections \ref{sec:devices} through \ref{sec:integration} then describe our approach: heterogeneous devices, error management, network architecture, and integration.

\section{The InterQnet Approach} \label{sec:approach}

Quantum communication networks must scale along multiple dimensions, including \textbf{heterogeneity} of qubit platforms,
\textbf{distance} between quantum information source and destination, \textbf{number of quantum nodes} in the network, \textbf{rate} of teleportation or entanglement distribution, and \textbf{number of users}.
As elaborated in \Cref{sec:open-challenges}, open scalability challenges include the integration of heterogeneous technologies (\textbf{C1}), the development of control frameworks (\textbf{C2}), and a refined understanding of architectures (\textbf{C3}) for quantum networks.
InterQnet takes a full-stack, co-design approach to enable scalable quantum networks through two overarching goals:
\begin{itemize}
    \item \textbf{InterQnet-Achieve:} Practical realization of a 3-node heterogeneous quantum network that integrates ytterbium  atoms, erbium  ions, and superconducting qubits, connected over the Argonne quantum network (ARQNET)~\cite{islam2024experiences} fiber infrastructure with quantum frequency conversion and microwave-optical transduction. InterQnet-Achieve aims to tackle challenge \textbf{C1} by conducting heterogeneous device development and integration, with experimental demonstrations of entanglement distribution. Moreover, InterQnet-Achieve will address challenge \textbf{C2} by adopting a centralized control architecture following SDN design principles.
    \item \textbf{InterQnet-Scale:} A systems study focused on architectural trade-offs critical to scalability, including centralized vs.~distributed control, in-band vs.~out-of-band signaling, and resource management across heterogeneous platforms. InterQnet-Scale aims to close the knowledge gap identified in \textbf{C3} by conducting modeling and simulation studies considering models informed by experimental data and large-scale network topologies.
\end{itemize}

These goals are supported by three tightly integrated technical thrusts: (1) Heterogeneous Devices, which develops and aligns physical-layer capabilities; (2) Error Management, which explores quantum error mitigation and correction strategies for communication; and (3) Network Architecture and Protocols, which designs and evaluates control and coordination across diverse topologies. 
Simulations are grounded in experimental measurements and conducted by using SeQUeNCe, a modular simulation framework developed by the team to support device-aware, protocol-driven architectural studies.

\begin{figure*}
    \centering
    \includegraphics[width=\textwidth]{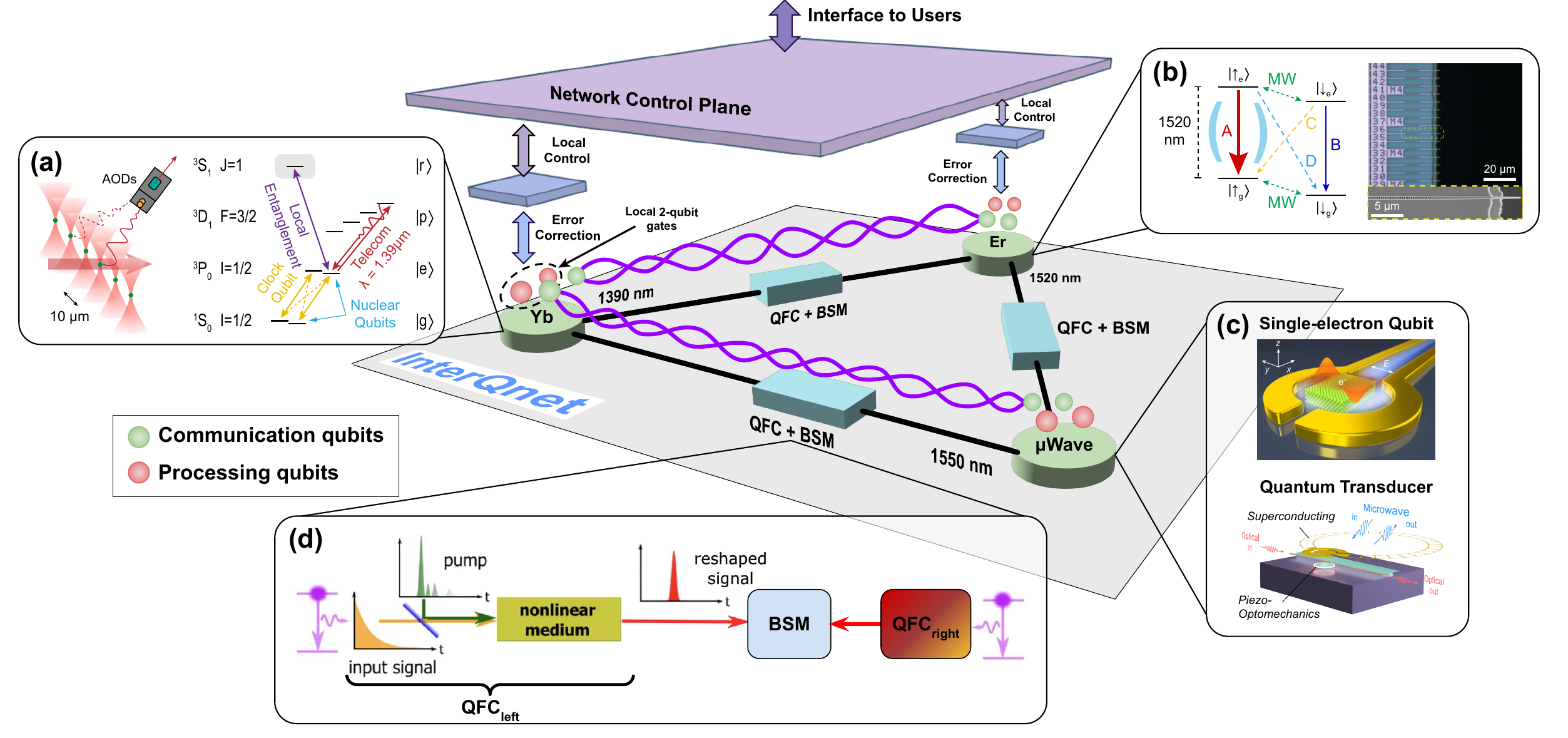}
    \caption{Three-node heterogeneous quantum network consisting of trapped ytterbium (Yb) atoms, erbium (Er$^{3+}$) ions in solids, and superconducting quantum devices ($\mu$Wave). BSM, the Bell state measurement,  includes a 50/50 beam splitter and a pair of single-photon detectors. (a) Array of trapped Yb atoms with the relevant level structure for controlling optical clock qubits and nuclear qubits. AOD is an acousto-optics deflector. (b) Energy-level structure of the Er$^{3+}$ qubit in a magnetic field with cavity-enhanced transition A. Also shown is an optical image of an array of photonic crystal cavities that host Er$^{3+}$ ions in a TiO$_2$ thin film on a silicon-on-insulator (SOI) wafer. (c) Schematics of the upper panel, a single-electron qubit trapped on a solid neon surface (green) strongly interacting with a superconducting resonator (yellow), and lower panel, an integrated quantum transducer consisting of a piezo-optomechanical microwheel resonator strongly coupled with a superconducting Ouroboros resonator. (d) Schematic of pulse shaping using a quantum frequency converter (QFC) followed by BSM. Table~\ref{tab:devices} summarizes the current development status of our devices.
    }
    \label{fig:triangle-qnet}
\end{figure*}

\section{Heterogeneous Devices} \label{sec:devices}
A quantum repeater is a key element in a scalable quantum network. Component devices such as quantum memories and quantum frequency converters are essential to the realization of a quantum repeater. 
We leverage three  types of state-of-the-art qubit systems that are being developed at Argonne as quantum nodes to demonstrate a quantum network repeater (\Cref{fig:triangle-qnet}) for InterQnet-Achieve: trapped ytterbium atom arrays, erbium ions (\Erppp) in solids, and superconducting quantum devices. This heterogeneous architecture addresses a key scalability requirement (see \textbf{C1} in \Cref{sec:open-challenges}) by allowing  remote quantum entanglement and communication between different kinds of quantum information carriers in both the optical and microwave regimes.

Each of the three qubit systems can generate either photon-photon or photon-spin entanglement to serve as a quantum node for entanglement swapping and entanglement-based quantum communications. 
Link-level entanglement between heterogeneous devices will be generated by using the meet-in-the-middle  protocol with QFC at a Bell state measurement (BSM) node.
We will use the Yb atom array system as the key quantum repeater station; this will provide both long-lived quantum memories based on the atom spin states and the necessary two-qubit gates between local atoms for implementing the error mitigation schemes of Thrust 2 (see \Cref{sec:qec}).
Below, we provide detailed descriptions of the three distinct qubit platforms and clarify the role of each component device.

\subsection{Ytterbium atoms}
Neutral atoms trapped in optical tweezer arrays \cite{norcia2018microscopic,cooper2018alkaline,saskin2019narrow,covey20192000} have become a compelling candidate for realizing key functions of a quantum repeater, such as high-fidelity, multiqubit entanglement distribution~\cite{madjarov2020high,omran2019generation}.
Yb atoms are of particular interest because their long-lived, optically excited $^{3}$P$_{0}$ ``clock'' state couples to several transitions in the telecom band (1390, 1480, and 1540~nm).  These transitions enable the generation of atom-photon entangled states where the emission time (or ``time bin'' of a telecom photon) is entangled with the atomic state. We recently demonstrated this atom-photon entanglement using the 1390~nm $^3$P$_0$ to $^3$D$_1$ transition in Yb~\cite{li2025parallelized}.   

As shown in \Cref{fig:triangle-qnet}(a), our work will initially focus on the $^{3}$P$_{0}$ -- $^{3}$D$_{1}$ transition at 1390~nm because of its relative simplicity \cite{covey2019telecom,li2025parallelized}.  Recent advances in visible-to-telecom quantum frequency conversion \cite{wengerowsky2025quantum} also make atom-photon entanglement mediated by visible transitions from the ground ($^1$S$_0$) state attractive alternatives owing to less complicated atomic state manipulation and less spontaneous emission to undesired states.

In order to achieve entanglement between two-matter qubits, the 1390~nm photons from a single atom are collected with a high numerical aperture microscope objective and sent through a long communication fiber to interfere with a photon emitted from another distant qubit system. After beam splitting and photon detection, a heralded Bell state (remote entanglement) can be generated between the atom and the distant qubit (e.g., another Yb atom, an Er ion, or a microwave qubit). 
With their long coherence time ($>10\,$s), Yb atoms can function as quantum memories to store the entanglement. 
Moreover, the atom array configuration will allow us to manipulate interactions between adjacent atoms, hence realizing quantum repeater operations.
For this, two Yb atoms would be first entangled with two distant nodes, and then local operations and measurements would swap the entanglement to the distant nodes. 
Combined with two-qubit gates, error mitigation schemes can be implemented to improve the fidelity of entanglement swapping and quantum communication.

\subsection{Erbium Ions} 
Erbium ions boast long spin~\cite{LeDantec2021,Rancic2018} and optical~\cite{Bottger2009} coherence times and exhibit a natural telecom-band optical transition near the 1520--1530~nm range.
These properties render them promising candidates for quantum memory applications.
Our work aims to leverage recent advancements in the nanophotonic integration of \Erppp\ to establish a \textit{telecom-wavelength solid-state quantum memory} node in ARQNET. Integration with nanophotonic cavities provides compelling advantages by significantly enhancing the single-photon emission rate via the Purcell enhancement and greatly increasing the scalability of erbium-based devices through CMOS-compatible nanofabrication. 

An optical image of the \Erppp\ device is shown in \Cref{fig:triangle-qnet}(b).
The device is a thin film of Er-doped titanium dioxide (TiO\textsubscript{2}) with low background spins, directly grown on top of a silicon-on-insulator (SOI) substrate.
Fabricating 1D nanophotonic crystal cavities to couple with the \Erppp\ results in an  optical photon emission  Purcell enhancement factor exceeding 200, as well as a high single-photon collection efficiency ($\sim$~40\%) via fiber-to-chip coupling~\cite{dibos2022NL}.
Similar devices of \Erppp\ in TiO\textsubscript{2} thin films grown post-fabrication on SiN waveguides boast a narrow ensemble homogeneous linewidth of 5~kHz and spectral diffusion of 27~kHz over 4~ms, well within the 1\,MHz microwave-optical (M-O) transducer bandwidth~\cite{gupta2025eqm}.

To entangle \Erppp\ with other qubits, we will use our cavity devices to selectively engineer a cycling transition~\cite{Raha2020,Chen2021}.
The initialization and preparation of a superposition spin state of the erbium electron can be accomplished by applying an excited state microwave (MW\textsubscript{e}) $\pi/2$-pulse followed by the emission of a telecom photon. 
Subsequently, a spin-selective optical $\pi/2$-pulse can be applied to rotate the spin state, thereby generating an entangled spin-photon state.

\subsection{Superconducting quantum devices} \label{sec:dev-sc}
A scalable quantum communication network should be able to not only transmit quantum information carried by optical photons but also interface with solid-state qubits working at the microwave frequencies~\cite{kimble2008quantum,Leon2021} for computation tasks. 
At the microwave quantum node, we will leverage GHz qubit devices based on superconducting transmons and/or single electron-on-solid neon (eNe) qubits \cite{zhou2022Nature} to \textit{create and receive microwave quantum communication signals} and integrated quantum transducers to \textit{generate M-O photon entanglement} \cite{Sahu2023,Meesala2024PRX}, hence achieving remote entanglement between a microwave qubit and a distant Yb atom or Er ion. 

As shown in the upper panel of \Cref{fig:triangle-qnet}(c), the eNe qubits feature a unique design by trapping isolated single electrons on an ultraclean, low-noise solid neon surface to achieve a record-long coherence time of $\sim$100\,$\mu$s with high-fidelity single-qubit gates ($>99.97\%$)~\cite{Zhou2023}, which outperform all existing semiconducting or superconducting charge qubits.
Recently, we have demonstrated remarkable noise resilience properties of the eNe qubits that allow qubit operations at elevated temperatures up to 0.4K~\cite{Li2025noise} as well as the coherent manipulation of multiple interacting eNe qubits~\cite{Li2025couple}.
In addition, state-of-the-art superconducting transmon qubits are available to prepare and communicate microwave quantum states, expanding the scalability of our heterogeneous quantum network architecture.
To interface these microwave qubits with telecom fibers, we will harness integrated superconducting piezo-optomechanical (POM) quantum transducers~\cite{han2020POM} (see \Cref{fig:triangle-qnet}(c) lower panel). 
The transducers utilize high-frequency phonons to mediate cavity-enhanced interaction between GHz microwave and 1550 nm optical photons to achieve a high M-O photon conversion efficiency with low noise~\cite{Han2021review}.
In this project we aim to substantially improve the transducer performance by developing ultra-high-Q POM resonators and reducing fiber-to-chip coupling loss and pump-induced heating, hence pushing the transducers into the quantum regime with a high efficiency of $>50$\% and subphoton added noise.

\subsection{Optical Quantum Frequency Converters} 
Quantum frequency converters (QFCs) provide the important capability to convert an input mode of light from one frequency to another while preserving its quantum state. A QFC was proposed and demonstrated by Kumar and coworkers \cite{Kumar1990OL, Huang1992PRL} in the 1990s and has since been recognized as a key component for emerging quantum networks.
In our heterogeneous network, different quantum nodes emit photons at different wavelengths (1390, 1520, 1550~nm for Yb, \Erppp, POM, respectively). 
Thus, we will develop high-efficiency QFC devices for \textit{converting single photons emitted from different qubit systems to a matched wavelength of 745.7~nm},  enabling two-photon interference and Bell state measurements to herald remote entanglement.

Key roadblocks to demonstrating high-efficiency, low-noise QFCs are noise processes, such as Raman scattering and weakly phase-matched parametric down-conversion, which arise in nonlinear media due to the strong pump laser field that is required to achieve high-efficiency frequency conversion of photonic quantum states via sum or difference frequency generation.  Recent advances in optical to telecom conversion have demonstrated that such noise can be dramatically reduced by limiting the output bandwidth of the QFC to a few megahertz since the noise processes are inherently broadband; however, overall device efficiency suffers because of a reliance on low transmission filtering elements \cite{wengerowsky2025quantum}.  

Our approach utilizes commercial ultra-narrow ($\sim100$ GHz), high transmission ($>95\%$), optical band-pass filters combined with modest finesse ($\mathcal{F}=200$) Fabry Perot optical cavities of multiple lengths to transmit only a single mode of the longest cavity.  The bandwidth of the transmitted light can be tuned to match a quantum networking device by properly selecting the length and finesse of the longest cavity.  For example, to convert 1389~nm light emitted from Yb atoms, a 10~cm cavity with $\mathcal{F}=200$ is used to transmit a bandwidth of $\sim7$~MHz, which is broader than the bandwidth of emitted light ($0.5$~MHz).  To maintain high transmission through the cavities, we utilize superpolished mirror substrates and low-loss optical coatings.

Besides bandwidth, conversion efficiency, and added noise, another important co-design parameter 
is the temporal shape of the single photons, which needs to be matched between different quantum emitters to remove photon distinguishability.
We have previously shown that such wavefunction engineering can be done by controlling the temporal shape of the pump beam \cite{Huang2013OL, Manurkar2017OL} during quantum frequency conversion. 
The capability of temporal mode shaping will play an important role in our heterogeneous network, allowing us to maximize visibility of two-photon interference and hence optimize Bell state measurements.

\begin{table*}[]
\centering
\caption{Summary of development status for heterogeneous devices.}
\label{tab:devices}
\begin{tabular}{|p{1.75cm}|p{1.5cm}|p{3.0cm}|p{3.0cm}|p{2.5cm}|p{3.5cm}|}
\hline
\textbf{Device} &
  \textbf{Application} &
  \textbf{State-of-the-Art} &
  \textbf{InterQnet-Achieve Goal} &
  \textbf{Status} &
  \textbf{Challenges} \\ \hline
Ytterbium Atom Tweezer Arrays &
   Entanglement distribution; Repeater; Error correction &
  \textless{}1~Hz spin-entangled photon emission rate, Fidelity \textgreater{}0.95 \cite{li2025parallelized}&
  1~Hz heralded entanglement rate, Fidelity \textgreater{}0.99 &
  Implementing spin-photon entanglement on ARQNET &
  Current system limited by photon collection efficiency. Coupling to optical cavities dramatically improves collection \cite{Huie2021}. \\ \hline
Erbium Ions in TiO$_2$ thin films &
  Memory &
  Spin-photon entanglement rate of $\sim$1.5~Hz over \textgreater{}15~km of optical fiber in hybrid single-ion devices &
  1~Hz heralded entanglement rate &
  Single-ion devices fabricated and tested in laboratory &
  Oxygen vacancies incorporated during film growth and induced during fabrication limit optical coherence \\ \hline
Superconducting and Single Electron Qubits &
  Computing; Sensing; Memory &
  Processors integrating $>$100 transmon qubits \cite{GoogleQuantumAI_2025_nature, Bravyi2024}, Coherence time $\sim$100 $\mu{s}$ \cite{GoogleQuantumAI_2025_nature, Zhou2023} &
  Superconducting and/or electron qubits with cavity frequency matched to M-O transducers in quantum network &
  Device design and initial fabrication completed &
  Frequency matching between qubit readout cavity and the M-O transducer \\ \hline
Microwave-optical (M-O) Transducer &
  M-O photon frequency conversion; Entanglement generation &
  Efficiency $\sim$10\% with added noise $\sim$0.2 \cite{Sahu2023} \par
  \textit{\scriptsize{(Higher efficiency of $\sim$48\% was achieved with higher noise and lower bandwidth \cite{Higginbotham2018})}} &
  High efficiency \textgreater{}50\% with low added noise $\lesssim0.1$ for quantum operations &
  Initial device fabrication completed. Device being characterized and will be further optimized &
  High optical pump power induces heating and noise at low temperature. Multiple frequency/wavelength matching between the transducer and other qubit systems \\ \hline
Optical Quantum Frequency Converters &
  O-O photon frequency conversion &
  Total device efficiency 25\%, negligible dark count noise \cite{wengerowsky2025quantum}&
  Total device efficiency \textgreater{}50\%, negligible dark count noise &
  Design completed, working toward lab demonstration &
  High-efficiency conversion creates noise that must be filtered or eliminated \\ \hline
\end{tabular}
\end{table*}

\subsection{Summary of Achievements and Open Challenges}
Table~\ref{tab:devices} shows a summary of the current development status of InterQnet devices.
For Yb atoms, we have demonstrated high-fidelity atom-photon entanglement directly in the telecom band---a key building block for remote entanglement. The main limitation today is photon collection efficiency. In the near term we are improving rates with parallel multiplexing; in the longer term we will couple atoms to optical cavities to enhance collection. This will bring us to the device performance parameter required for InterQnet-Achieve, nameky, 1~Hz heralded entanglement rate with fidelity \textgreater{}0.99.
In Er-doped TiO\textsubscript{2} we have achieved a $1000\times$ narrowing of homogeneous linewidths at sub-Kelvin temperatures. This is critical for isolating single ions and achieving long coherence. Next steps include fabricating high-Q cavities and verifying performance with single emitters for telecom-compatible quantum memory to achieve heralded entanglement rates of 1~Hz.
For superconducting devices, we are nearing completion of two key components: frequency-tunable qubit readout resonators and high-efficiency, low-noise piezo-optomechanical transducers. Upcoming work will focus on engineering the nonlinearity and tunability of the readout resonator and boosting mechanical Q to improve the photon conversion efficiency of the transducer. These efforts will get us closer to the required \textgreater{}50\% conversion efficiency with low added noise ($\lesssim0.1$) for quantum operations for the InterQnet-Achieve demonstration.
In frequency conversion, we have recently completed classical characterization of the devices, and we are now investigating noise sources---especially Raman noise---to support quantum-compatible conversion (i.e., efficiency \textgreater{}50\% with negligible dark count noise).
Altogether, these efforts represent diverse but converging progress toward scalable, hardware-aware entanglement distribution across heterogeneous platforms.

To assist in the design and evaluation process, we are creating models for the above-mentioned key quantum devices in SeQUeNCe and applying numerical optimization methods to search for critical device parameters to maximize the fidelity of simulated networks. 
We have successfully modeled the quantum transducer, Yb atom node, and QFC in SeQUeNCe~\cite{laura2025transducer,miller2025simulationheterogeneousquantumnetwork}; the \Erppp\ device is work in progress.
Via simulations, we will not only better understand the behaviors of current experiments but also obtain insights for future device designs and improvements.
We will conduct ``what-if'' studies that will help capability development in future devices. For example, \textit{what if the dark count in the detectors is reduced by a factor of 10, or atomic operations can run an order of magnitude faster?} 
Furthermore, we will develop specialized numerical optimization methods to optimize operation conditions for high-fidelity quantum communications in future large-scale quantum networks.

\section{Error Management} \label{sec:qec}
Different from qubit-based quantum computing, optical quantum communication relies on bosonic modes to generate entangled states for long-distance transmission, without the need for universal gate sets. As a result, the quantum circuits involved in error management methods that can tolerate photon loss, imperfect entanglement generation, and heterogeneous device noise are usually simple and practical for near-term testbeds and scalable to future large networks. 
To aid in addressing challenges \textbf{C2} and \textbf{C3},
InterQnet adopts a layered approach to error management: (1) lightweight error detection via Pauli check sandwiching (PCS) and postselection, (2) resource-aware strategies that leverage gauge freedom to simplify correction, (3) entanglement distillation protocols that convert multiple noisy Bell pairs into fewer high-fidelity ones, and (4) advanced quantum LDPC (qLDPC) and bosonic codes that enable scalable, fault-tolerant entanglement purification. 
Hardware-level realizations of these approaches have been partially demonstrated or proposed over the past decades. An early work \cite{Reichle2006} demonstrated primitive entanglement purification via postselection. Although quantum error-correcting codes (QECCs) have not been widely used in entanglement distillation, promising schemes that apply more scalable and powerful quantum codes have recently been proposed for superconducting qubits \cite{GoogleQuantumAI_2025_nature, eickbusch2025demonstratingdynamicsurfacecodes} as well as for trapped ions and atom arrays \cite{PRXQuantum.5.030326, Bluvstein2026, Xu2024, Bravyi2024}. For bosonic quantum systems, error-correction protocols have been experimentally demonstrated in \cite{Ofek2016, Ni2023, Sivak2023, deNeeve2022, Putterman2025}. 
Together, these techniques form a hierarchy that links the device-level realities of \Cref{sec:devices} to the architectural protocols of \Cref{sec:net-arch}, and they map naturally onto the two project goals: enabling practical demonstrations in InterQnet-Achieve and guiding at-scale design choices in InterQnet-Scale.

\subsection{Pauli Check Sandwiching}

\begin{figure*}
    \centering
    \includegraphics[width=0.99\textwidth]{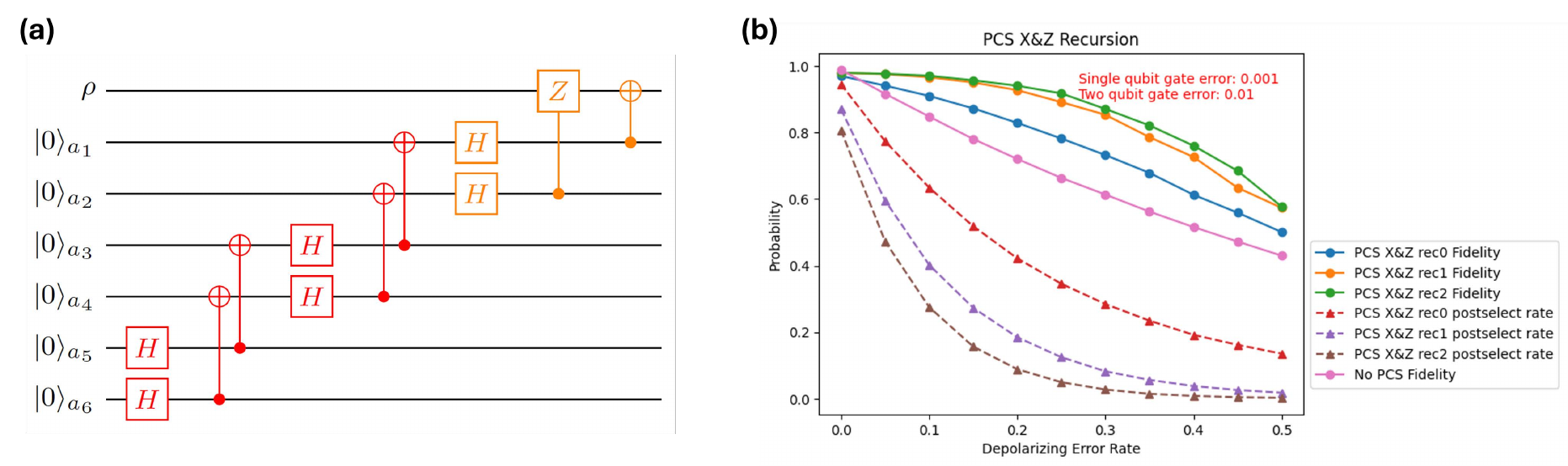}
	\caption{(a) Pauli checks forming distance-2 code. Red checks can be applied recursively. (b) Noisy simulation for different levels of recursion: rec0 represents no red gates; rec1 represents adding ancillas $a_3$ and $a_4$ and the supported red gates to the rec0 circuit; rec2 corresponds to adding ancillas $a_5$ and $a_6$ and the supported red gates to the rec1 circuit.}
	\label{fig:npcs}
\end{figure*}

In previous work we investigated quantum error detection based on Pauli check sandwiching (PCS)~\cite{Gonzales_2023PCS}, which shares many properties with quantum codes. PCS uses left and right Pauli checks to detect errors. It provides protection on the targeted qubits and generally incurs less resource overhead than standard quantum error correction and detection codes do. From the point of view of quantum codes, PCS is generally a distance 1 code (i.e., it cannot detect arbitrary single-qubit errors). Additionally, there is a direct correspondence between quantum codes and entanglement purification, since any qubit stabilizer codes can be converted to a one-way entanglement purification protocol due to the simulability of teleportation, and vice versa~\cite{Aschauer_2005QuantCommInNoisyEnv, Dur_EntPurAndQEC}. Thus, the performance of PCS in quantum repeater networks is an important area of investigation for InterQnet-Scale.

The PCS scheme~\cite{Gonzales_2023PCS} for quantum networks performs PCS in a distributed fashion, where the left checks are performed by the origin node and the right checks are performed by the receiving node (see Alg.~\ref{alg:npcs_pcs} for details). The left and right checks are chosen from the $+1$ elements of the Pauli group and must multiply to identity.

\begin{algorithm}[t]
\caption{Network PCS for distributing a Bell pair}
\label{alg:npcs_pcs}
\begin{algorithmic}[1]
\Require Nodes $A,B$; Pauli checks $\{P_j\}_{j=1}^m$; network channel $\mathcal{N}_{A\to B}$
\Ensure \textsc{Accept} a (postselected) shared Bell pair on $(A,B)$ or \textsc{Reject}
\State $A$ prepares $|\Phi^+\rangle_{A\,F}$ \Comment{$F$ is the half to be sent to $B$}
\For{$j=1$ to $m$}\Comment{left check(s) at node $A$}
  \State $a_j \leftarrow |+\rangle$ \Comment{check ancilla prepared at $A$}
  \State apply controlled-$P_j$ with control $a_j$ and target $F$ 
\EndFor
\State send $(F,\{a_j\}_{j=1}^m)$ through the network:
      $(F,\{a_j\}) \xrightarrow{\;\mathcal{N}_{A\to B}\;} (B,\{a_j\})$
\For{$j=m$ to $1$}\Comment{right check(s) at node $B$}
  \State apply controlled-$P_j$ with control $a_j$ and target $B$ 
  \State measure $a_j$ in $X$ to obtain $s_j \in \{0,1\}$
\EndFor
\If{$\forall j:\ s_j = 0$} \State \Return \textsc{Accept} \Else \State \Return \textsc{Reject} \EndIf
\end{algorithmic}
\end{algorithm}

The novel contributions of our current work~\cite{Gonzales_2025DetectingErrsInAQuantNetWithPauliChecks} are as follows. Assuming depolarizing errors, we derived the analytical equations for the postselected fidelity and postselection rate as a function of the noise for PCS $X$ checks and PCS $X\&Z$ checks. For PCS $X$ checks we have
\begin{align}
    c=\frac{1}{9}(1+2F)^2
\end{align}
and
\begin{align}
    F'=\frac{9F^2}{(1+2F)^2},
\end{align}
and for PCS $X\&Z$ we have
\begin{align}
    c=\frac{1}{324}(3+6F-\sqrt{12F-3}+4F\sqrt{12F-3})^2
\end{align}
and
\begin{align}\label{eq:pcsxz_fid}
    F'=\frac{1+52F^2-\sqrt{12F-3}-2F(4+\sqrt{12F-3})}{(\sqrt{12F-3}-1-8F)^2},
\end{align}
where $c$ is the postselection probability, $F'$ is the postselected Bell state fidelity, and $F$ is the initial fidelity (i.e., fidelity of the Bell state without PCS). Next, we simulated our scheme assuming depolarization noise on the optical fiber channel and witnessed fidelity gains. The fidelity gains also appeared in simulations with noisy gates. PCS can be enhanced by adding $Z$ checks on the ancilla qubits. We can apply this process recursively by also checking the new ancillas with $Z$ checks. We  discovered that this process generates a family of distance 2 codes. The structure of this family is described in \Cref{fig:npcs}(a), where the red gates can be applied recursively. Since the code distance is fixed, only the first recursion of ancilla checks provides significant improvements in fidelity. We also simulated this method with noisy gates and attained fidelity improvements as shown in \Cref{fig:npcs}(b). We vary along the x-axis the depolarization error rate of the flying qubits. The baseline without PCS is in pink. The blue, orange, and green lines are no recursion, one recursion, and two recursions, respectively. The dashed lines represent postselection rates. Since the code distance is fixed, the green and orange lines yield similar fidelities.

\subsection{Quantum Resource Correction}
In certain cases we may be interested only in correcting certain properties of a quantum state. Intuitively, this leads to less resource overhead than standard QECCs, which aim to correct the logical state exactly. 
In quantum resource theory~\cite{Chitambar_2019}, 
restricted control defines a set of free operations $\mathcal{O}$ and free states $\mathcal{F}$. States outside $\mathcal{F}$ possess a quantum resource. The \textbf{amount of resource} in a state quantifies the degree to which it deviates from the free set, and thus how useful it is for tasks that cannot be accomplished by using only free states and operations. By definition, free states have zero resource, and the amount of resource cannot increase under free operations. This general notion underlies familiar quantities in quantum information such as entanglement, coherence, asymmetry, and nonlocality.
Thus, in~\cite{byrd_2025quantumresourcecorrection} we investigated error-correction schemes that restore a specified quantum resource while allowing other properties of the logical state to change (i.e., correcting the amount of a quantum resource).

In quantum networks, entanglement is the most commonly used resource. 
We derived necessary and sufficient conditions for correcting the amount of a logical resource in a logical state. The resources considered are those that are left invariant by a group of unitary operations (e.g., local unitaries for entanglement).
While the conditions are equivalent to the QECC conditions, recovery is simplified. Resource-preserving operations can result in gauge freedom in the code space. For instance, for the $l_1$ norm measure of coherence~\cite{Baumgratz_2014QuantCoherence} and incoherent basis in the standard logical basis, logical Pauli operations are gauge transformations. In contrast, in standard QECC, unwanted logical Pauli operations are logical errors. Thus, the optimization in decoders for stabilizer codes (such as the surface code) is simpler for resource correction than in QECC. This is an important result since the general decoding problem is in the class of NP-complete or \#P-complete problems, depending on whether degenerate decoding is used  \cite{deMarti_iOlius_2024DecodingAlgForSurfCodes}. 
Importantly, fast decoders can have the same accuracy as generally more accurate decoders when applied to resource correction. We also investigated extending the logical incoherent basis over the physical space and identified channels that preserve the coherence of any state in the code space. Consequently, stabilizer measurements can be combined in many situations. 

\subsection{Entanglement Distillation and Error Detection/Correction} \label{sec:qec-dual-species}
In principle, entanglement distillation can asymptotically extract multiple copies of perfect Bell pairs, limited only by the distillable entanglement~\cite{Bennett96c, AJOSS2025}. 
In practice, with finite noisy entangled inputs, the majority of purification protocols reduce noise and concentrate entanglement in the output but do not yield perfect Bell pairs. For instance, certain protocols require four noisy states as input to suppress depolarizing errors in the output state effectively. From an engineering perspective, these improved entangled states from finite-size protocols are valuable for constructing nontrivial applications.

By exploiting the correspondence between quantum error correction/detection and entanglement distillation \cite{Aschauer04, Aschauer_2005QuantCommInNoisyEnv, Dur_EntPurAndQEC}, we can devise a wide array of efficient entanglement purification protocols, assisted by one-way or two-way classical communication, based on quantum error-correcting codes. For example, \Cref{fig:QEC}(a) depicts a quantum circuit based on the $[[4,2,2]]$ code, also known to be equivalent to the Leung--Shor protocol~\cite{Leung_Shor_2008}. Upon success, it transforms four noisy inputs into two purified entangled states with higher fidelity.

To bridge the gap between theory and practice, we analyzed the circuit implementations of entanglement purification on popular platforms (e.g., neutral atom systems) in~\cite{li2025efficiententanglementpurificationcircuit}. 
Real physical systems often impose additional constraints that hinder the efficient execution of the desired protocols. In Rydberg atom array systems, leveraging their unique features, we demonstrated the robustness and universality of efficient entanglement purification protocol implementations.

This advancement enhances our understanding of reliably establishing entanglement over long distances. It also facilitates the development of advanced quantum repeater architectures across various generations~\cite{Muralidharan14, Munro15, Muralidharan16b}. We are conducting simulations and evaluating novel protocols using SeQUeNCe, the Stim~\cite{gidney2021stim} Python package, and QETLAB~\cite{qetlab} to understand the scalability trade-off of deploying these protocols in quantum networks.

\subsection{qLDPC-Based Entanglement Distillation} \label{sec:qldpc}
\begin{figure}[t]
   \centering
   \includegraphics[width=\columnwidth]{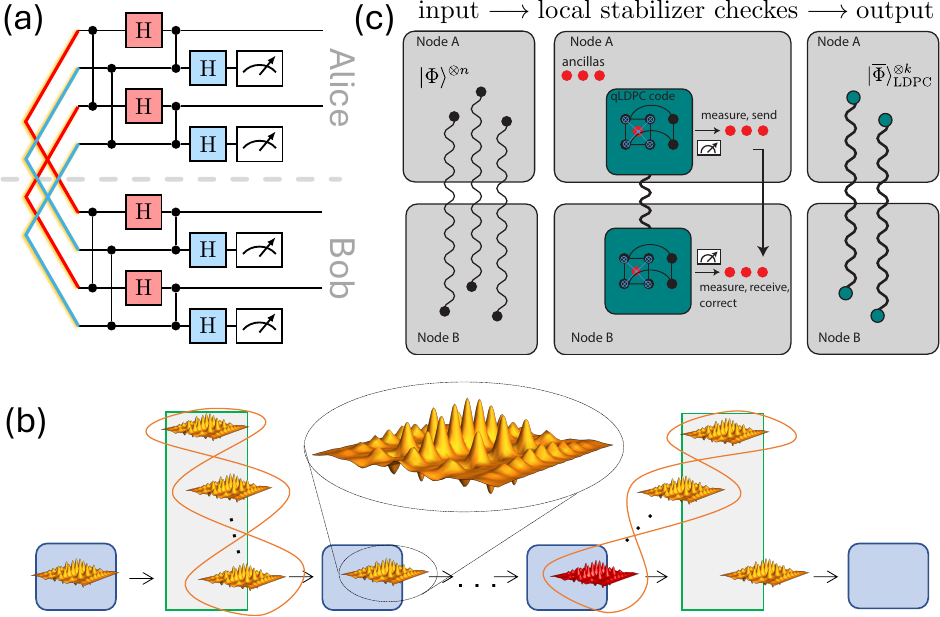}
   \caption{(a) Quantum error detection for novel entanglement distillation protocol that can distill two Bell pairs from four imperfect ones. (b) Quantum error correction with concatenated bosonic GKP code and qLDPC code. (c) Efficient and robust purification of logical Bell pairs with qLDPC codes.}
   \label{fig:QEC}
\end{figure}

We are designing robust bosonic quantum error-correcting codes specifically tailored for quantum communications, which rely on optical wave packets for communication over long distances. 
We aim to enhance the performance of quantum error correction by optimizing both the choice of the logical subspace and the recovery operation. Previous investigations have shown that the single-mode Gottesman--Kitaev--Preskill (GKP) code achieves optimal performance for encoding with a single bosonic mode, effectively mitigating dominant excitation loss errors in quantum communication~\cite{Noh18}. Furthermore, a recent experiment has demonstrated the above-break-even performance of the GKP code~\cite{Sivak23}, which is highly encouraging.
However, open questions remain regarding the exact ultimate performance limit and the achievement of optimized recovery, even for single-mode GKP encoding: \textit{Can we restore the encoded quantum information with fidelity above 99.99\% even with a 10\% loss rate?} Moreover, extending this approach practically to multiple bosonic modes to approach the quantum channel capacity remains unclear. To address these challenges, our approach involves two key steps. (1) We will use the transpose-recovery-channel method (known as Petz recovery map) to explicitly construct a recovery operation that is close to optimal for the GKP code. This will not only provide an asymptotic performance limit for single-mode GKP encoding but also improve the design of the experimental GKP-QEC recovery operation. (2) We will investigate the concatenation of bosonic quantum codes with the recent development of good quantum low-density parity-check codes~\cite{Panteleev21, Leverrier22}, which exhibit favorable scaling in both coding rate and distance rate (see \Cref{fig:QEC}(b)). This will enable the development of multimode encoding schemes that approach the quantum channel capacity for future quantum networks~\cite{Weedbrook12}.

Recent progress in entanglement purification involves the use of various quantum low-density parity-check (qLDPC) codes~\cite{ataides2025}. \Cref{fig:QEC}(c) illustrates the purification of Bell states at the logical level. The low density of stabilizer checks mitigates undesirable error propagation. In simulations of noisy quantum circuits, as the system size increases, we observe improved purified entanglement resources. Its scalability and low overhead of encoding and decoding render it an ideal candidate subroutine for future quantum networks.

\subsection{Summary of Achievements and Open Challenges}
We have investigated several complementary approaches to error management. Pauli check sandwiching  improves fidelity through Pauli operator checks and postselection, although implementing the necessary entangling gates remains challenging. Quantum resource correction (QRC) exploits gauge freedom to construct resource-preserving recovery maps that simplify decoding relative to conventional QECC. Entanglement distillation protocols, designed for dual-species atom arrays and compiled into circuits with only global control, enable purification on hardware with limited addressability, although coupling and measurement performance must improve. At a more advanced level, quantum LDPC codes leverage sparse stabilizers to achieve fault-tolerant purification with high thresholds, while multimode bosonic codes such as GKP, combined with Petz recovery, can approach channel capacity~\cite{Zheng2025PRXQ, Li2025PRL}; concatenating these methods provides a scalable path toward efficient entanglement distribution.

Ongoing work is expanding these methods to more general noise models and developing collective bosonic encodings, offering a range of scalable, hardware-aware solutions for robust entanglement distribution. A key near-term goal is to demonstrate improved Bell violations on ARQNET using PCS. Violating a Bell inequality marks a transition away from local hidden-variable models and provides a clear experimental milestone for InterQnet-Achieve.

Together, these approaches provide a layered strategy. Lightweight PCS, QRC, and small-code distillation enable practical demonstrations in InterQnet-Achieve, while advanced qLDPC and bosonic codes inform scalability analyses in InterQnet-Scale.
To elaborate on the technical feasibility of employing these methods in ARQNET, we focus our discussion on atom and superconducting qubit platforms (see also \Cref{tab:devices}).
Multiqubit control has been achieved by all these platforms with >99\% fidelity, sufficient for QEC demonstrations~\cite{GoogleQuantumAI_2025_nature,Bluvstein2026}.
Quantum memory lifetimes (compared with typical classical communication time, millisecond or longer) impose an interesting trade-off.
For atom platforms, the memory lifetime can be very long (approaching/exceeding a second), while the memory lifetime for superconducting qubits is often limited (approaching milliseconds for transmon qubits).
Comparing these memory lifetimes with entanglement generation rates, atom platforms can achieve generation rates $\sim$1~Hz~\cite{li2025parallelized},  close to the memory decoherence rate.
For superconducting platforms, we can perform pitch-and-catch microwave excitations to generate entanglement between different dilution refrigerators (over 30 meters~\cite{storz2023loophole}). Hence, with good transmon qubits, we should be able to demonstrate entanglement distillation. 
For long-range entanglement generation, we will need M-O transduction before we can generate long-range entanglement of superconducting qubits. This is an active research area, as  discussed in \Cref{sec:dev-sc}.

\section{Network Architecture and Protocols} \label{sec:net-arch}

\begin{figure}
    \centering
    \includegraphics[width=\columnwidth]{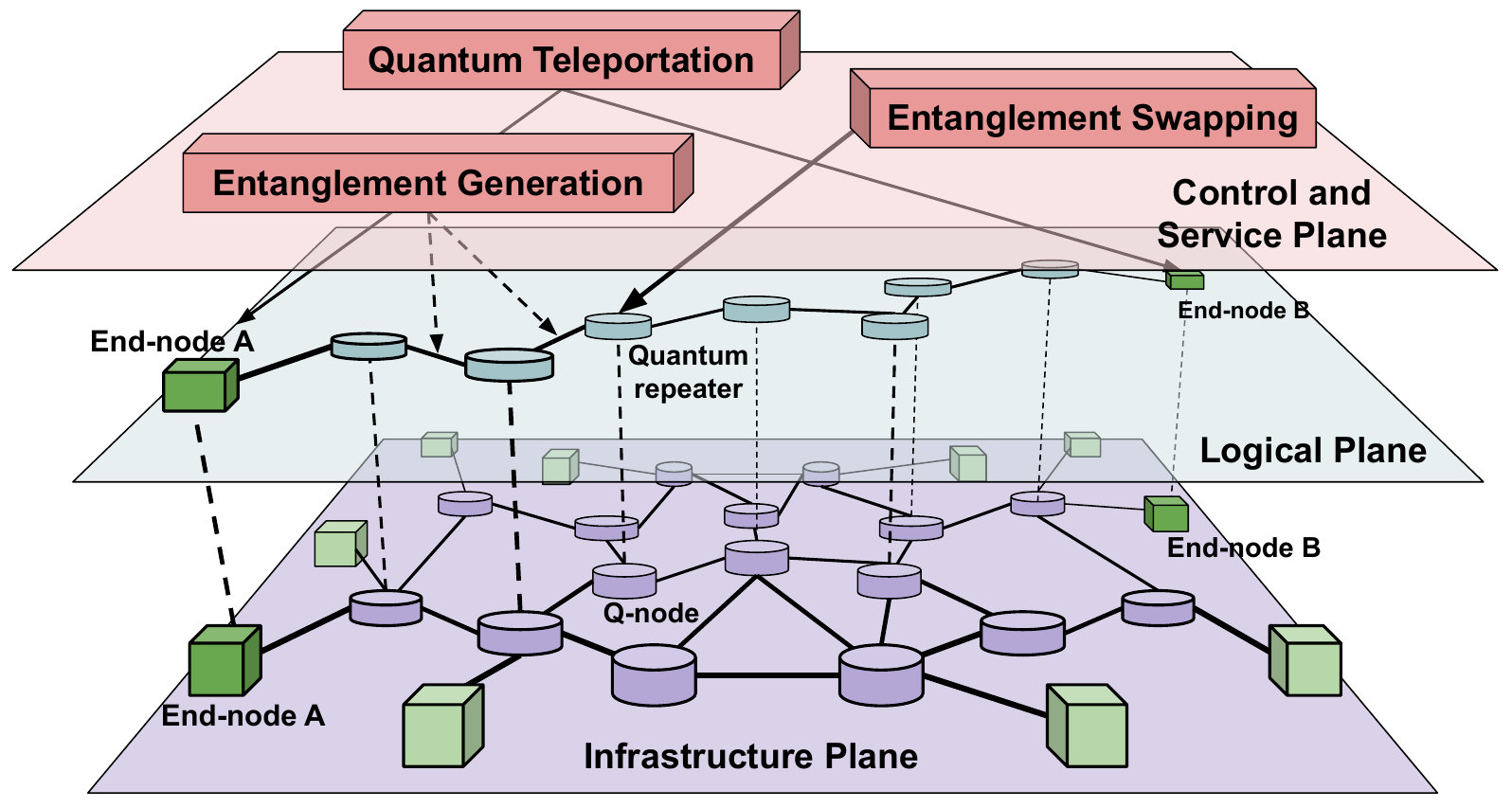}
    \caption{Architecture of a scalable quantum network using a plane abstraction.}
    \label{fig:interqnet-arch}
\end{figure}

This thrust directly tackles challenge \textbf{C3} by conducting systems studies focused on architectural trade-offs for InterQnet-Scale.
\Cref{fig:interqnet-arch} illustrates a scalable quantum network organized using a three-plane abstraction. The infrastructure plane at the bottom represents all physical devices and fiber connections. The logical plane in the middle captures the repeater-chain abstraction once a network orchestrator has identified a suitable path between end nodes (e.g., A and B in dark green). The control and service plane at the top encompasses orchestration, routing, and user-facing services accessed through standard interfaces. We adopt this layered abstraction to structure our work on network architecture and protocols, enabling systematic study of scalability across distance, node and user counts, heterogeneity, while providing a framework for integrating device models from \Cref{sec:devices} and error management methods from \Cref{sec:qec} into protocol-level evaluations.

\subsection{Faithfully Simulating the Infrastructure Plane of Quantum Networks}
Given the heterogeneous nature of InterQnet, SeQUeNCe has been designed to work with multiple architectural paradigms of the infrastructure plane. 
For this iteration of the work, we report simulation modes to tackle two important aspects of quantum communications: optical and atomic communications. 
The following subsections present updates on the development of software extensions to SeQUeNCe that will enable future large-scale studies.

\subsubsection{Simulating Optical Quantum Communications}
The objective of simulating optical quantum communications is to analyze the effect of noise on realistic communication experiments. However, modeling realistic systems requires simulating large Hilbert spaces. Density matrices soon become computationally intractable to model large systems. To address this issue, we use tensor networks~\cite{orus2014practical} to simulate the sparsely entangled quantum states in quantum communications. 
Moreover, to model noise using tensor networks, we implement trajectory-based simulations to simulate open system dynamics.

\begin{figure}
    \centering
    \includegraphics[width=0.99\linewidth]{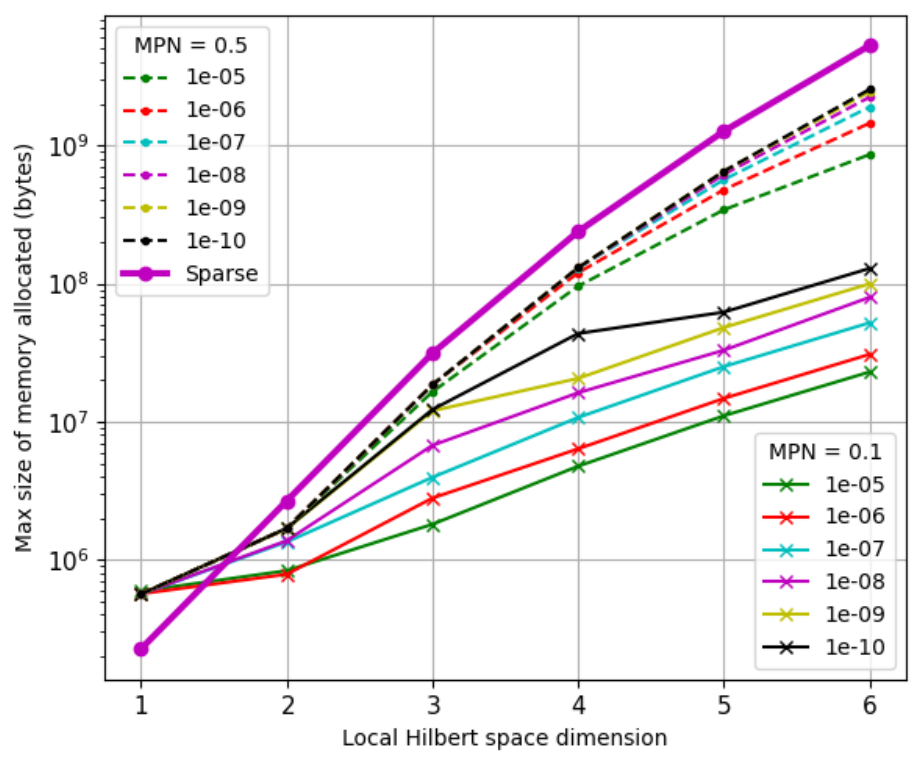}
    \caption{Simulation memory vs. local Hilbert space dimension. The simulation was performed over multiple cut-off values of tensor network decomposition errors for two different mean photon numbers (MPN).}
    \label{fig:TN_memory}
\end{figure}

Figure~\ref{fig:TN_memory} compares the memory requirements for simulating a polarization-based entanglement swapping experiment~\cite{scherer2009quantum} using tensor networks versus sparse density matrices, showing a substantial reduction in memory usage as the Hilbert space dimension increases. 

Kraus operators corresponding to the relevant noise models (e.g., amplitude damping, amplification, and depolarization) are applied to the matrix product states as they evolve through the system. These Kraus operators are applied probabilistically, with probabilities determined by the unnormalized weights of each trajectory \cite{isakov2021simulations}. Computing these probabilities for every iteration of a Monte Carlo experiment is computationally expensive and inefficient, thus we developed a novel trajectory caching scheme. Since most quantum channels in communication systems have certain trajectories that occur far more frequently than others, the required calculations for these common trajectories are performed only once, after which the cached results are reused for the majority of Monte Carlo samples.

\subsubsection{Simulating Physical Qubits and Their Emission}
Another challenge of faithful quantum network simulation lies in simulating entanglement generation protocols using a realistic hardware model of physical qubits. The fidelity and success rate of a protocol for generating entanglement depend on the hardware specifications, which consist of two distant qubits connected via an optical fiber link. A realistic simulation is essential for accurately predicting results that are comparable to experimental outcomes.
We constructed a model in the Lindblad form to account for the internal dynamics of physical qubits and photon emission into optical fibers, as well as decoherence and imperfections. We are developing a module that enables SeQUeNCe users to precompute the outcomes of entanglement generation protocols for use in the network simulation. The module aims to provide sufficient flexibility for various user-defined physical qubits and operational sequences of protocols.
We are also exploring possible ways to integrate this physical qubit simulation with the fiber-optical simulation previously described.

\subsubsection{Simulating a Heterogeneous Quantum Network} \label{sec:sim-het-net}

To help understand engineering challenges in integrating heterogeneous nodes in a quantum network, we have developed in~\cite{miller2025simulationheterogeneousquantumnetwork} faithful models of Yb-atom and superconducting nodes, as well as QFC and quantum transducers. In this preliminary study, we implemented entanglement generation and swapping protocols for time-bin encoded photons that account for disparate clock rates, quantum frequency conversion, and transducer losses/noise introduced by the heterogeneity. Using extensive simulations, we mapped the rate–fidelity trade space and identified the dominant bottlenecks unique to heterogeneous systems.

\begin{figure}
    \centering
    \includegraphics[width=\linewidth]{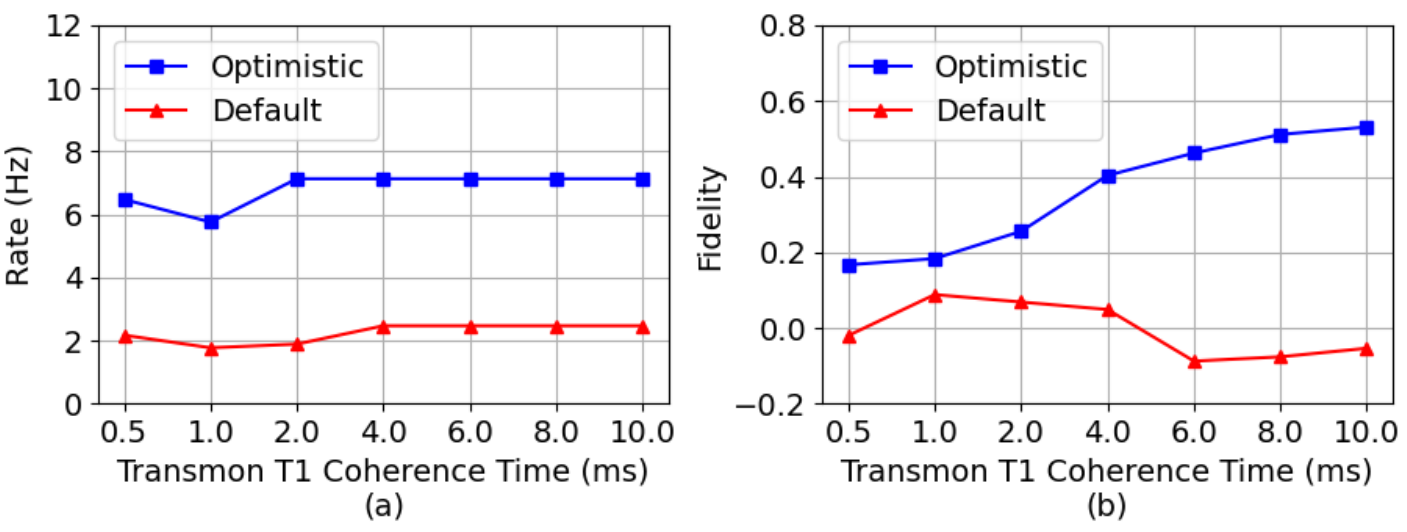}
    \vspace{-0.3in}
    \caption{Simulation results for entanglement generation between two superconducting nodes connected via an Yb-atom repeater, reproduced from~\cite{miller2025simulationheterogeneousquantumnetwork}. We vary the coherence time of the superconducting node and compare the entanglement rate (a) and fidelity (b) for both currently achievable and optimal device parameters in SeQUeNCe.}
    \vspace{-0.2in}
    \label{fig:uw-yb-uw}
\end{figure}

Figure~\ref{fig:uw-yb-uw} shows the simulation results for entanglement generation between two superconducting nodes connected via a Yb-atom repeater. We vary the transmon's $T_1$ coherence time from its default of $0.5~ms$ to a maximum of $10~ms$. For the optimal parameters (blue lines in the plots), all efficiency parameters are set to $100\%$, and noise rates are $0\%$. In contrast, the default device parameters assume near-term optimistic values, such as a 50\% collection efficiency for Yb and a 60\% transducer efficiency (see Table~I in~\cite{miller2025simulationheterogeneousquantumnetwork}). As expected, the rate with optimal parameters ($\sim 7~Hz$) is higher than with the default ($\sim 2~Hz$). We observe that the fidelity remains close to $0$ for default parameters, while an increase in coherence time leads to tangible increases in fidelity (up to $>0.5$ at $T_1 =10ms$).

\subsection{Entanglement witnessing in the logical plane of quantum networks}
Resilience is typically defined as the ability to survive either a malicious or unintentional network failure.
To fully ensure resilience in the logical plane of scalable quantum networks, one must ascertain that the distributed entangled states are actually shared between the desired users.  
For instance, noise could deteriorate the entanglement held in quantum memories of a repeater before a swap operation is scheduled.
Or more dramatically, two end nodes may have established entanglement with a malicious repeater~\cite{satoh2021attacking}.
In both cases, we want the entanglement distribution schemes to be resilient and amenable to verification.  
Therefore, we consider the problem of efficiently witnessing or certifying two or more sources of entanglement in a network by performing joint measurements instead of just single-qubit measurements.
By using joint measurement at network nodes, it becomes possible to simultaneously certify the distribution of entanglement from multiple sources using fewer shots than certifying the sources independently.

Our solution is to use tools from quantum self-testing and semi-device-independent quantum information processing to develop entanglement witnessing tests.  Similar to how the famous CHSH inequality can certify that entanglement has been distributed to two distant locations, our work will develop similar inequalities or tests that can certify the entanglement of two or more sources on a network.
Prior work has already shown the utility of this approach in special scenarios~\cite{Doolittle-2023a}, but the general methodology applied to simultaneous entanglement witnessing is relatively unexplored.
We anticipate making new progress by invoking prior results in the study of local operations and classical communications to help develop entanglement witnesses.
One key initial accomplishment is the development of a certification test that can simultaneously witness two noisy entangled states that are distributed in a three-node line network, known as the bilocal network.  
While this result is not significantly different from prior known results~\cite{Branciard-2012a}, it is a starting point for this research.  
The biggest challenge is that the optimization problems we encounter are non-convex, and therefore standard numerical techniques are not efficient.  We are exploring certain relaxations that can make the numerical optimizations more tractable.

\subsection{Architectural Approaches to the Control and Service Plane of Scalable Quantum Networks}
Here we study the trade-off of generating entanglement on demand vs. continuously generating link-level entanglement, with the goal of reducing the time to serve user requests.
In prior work~\cite{kolar2022acp} we proposed a theoretical framework that enables the continuous generation of entangled pairs between neighboring nodes, similar to \cite{talsma2024continuously} and \cite{gu2023esdi}.
This approach can adapt to incoming client requests to reduce the time to serve future requests.
In~\cite{kolar2022acp}, however, the theoretical framework is evaluated only through small-scale, ad hoc, and na\"ive time-slotted simulations.
The challenge is to implement the theoretical framework into a comprehensive quantum network protocol and simulate the protocol in a discrete-event quantum network simulator at a large scale, including realistic physical parameters.

The challenges were addressed in~\cite{zhan2025acp}, where we designed and implemented the Adaptive, Continuous entanglement generation Protocol (ACP).
ACP can reduce the time to serve a client request by letting a node attempt to generate entangled pairs with neighboring nodes in the background continuously, regardless of incoming requests.
The selection of neighbor nodes adapts to the request traffic by a feedback control algorithm.
Additionally, entanglement distillation is used to enhance the fidelity of entangled pairs.
We extended SeQUeNCe to support the simulation of ACP.
In particular, we added a single-heralded entanglement generation protocol that accounts for pregenerated entangled pairs.

\begin{figure}
    \centering
    \includegraphics[width=0.99\linewidth]{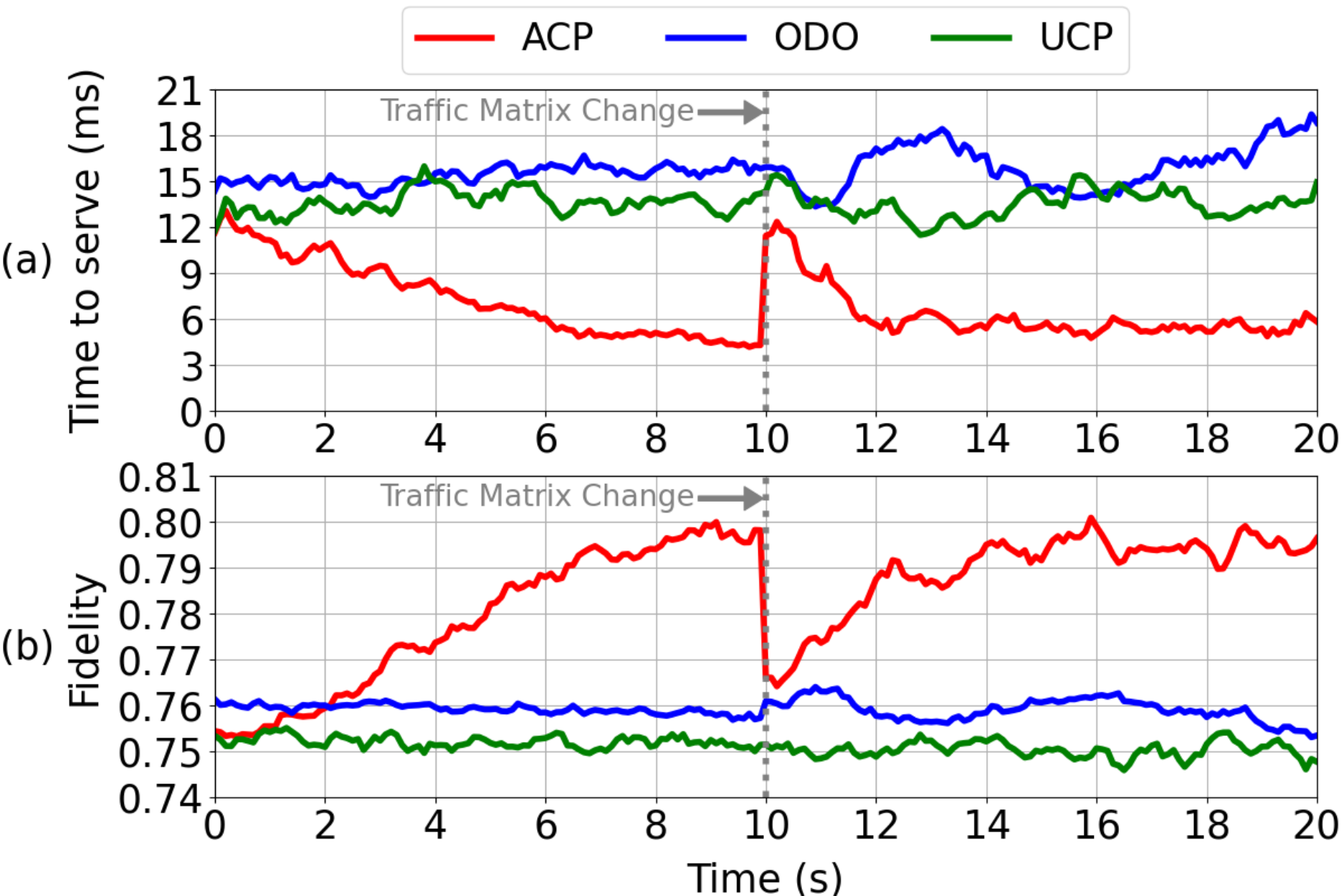}
    \caption{Simulation results for the 200-node autonomous system topology network, plots reproduced from~\cite{zhan2025acp}. (a) ACP outperforms on-demand-only (ODO) and uniform-continuous protocol (UCP) policies by having a 50+\% lower time to serve than UCP. (b) ACP enhanced with entanglement distillation can increase the fidelity of the end-to-end entangled pairs by as much as 0.05 compared with UCP and 0.04 compared with ODO, both without entanglement distillation. Both plots show the adaptiveness as the traffic pattern changes in the middle of the simulation.
    }
    \label{fig:acp-200}
\end{figure}

In~\cite{zhan2025acp}, we showed promising results when we evaluated the ACP in the SeQUeNCe simulator under various scales (2-node, 20-node, and 200-node networks) and realistic physical parameters. 
Figure~\ref{fig:acp-200} shows selected results from our prior work.
Although we assumed homogeneous quantum memories, ACP can reduce the time to serve by up to 94\% and improve fidelity by 0.05 in large-scale simulations.
We are now working on integrating the models described in \Cref{sec:sim-het-net} and ACP to simulate large-scale heterogeneous quantum networks.

\subsection{Summary of Accomplishments and Open Challenges}

We have made significant progress in creating faithful and computationally efficient simulations of the infrastructure plane of quantum networks. 
We developed simulation tools based on tensor networks to simulate optical communications in multidimensional bosonic states efficiently. These tools will serve as the foundation to evaluate the qLDPC entanglement distillation protocols described in \Cref{sec:qldpc}.
Our realistic implementation of an adaptive strategy to continuously generate link-level entanglement significantly outperforms the on-demand-only strategy by reducing the time to serve by up to 94\% and by improving fidelity by 0.05 in large-scale simulations.
We are now scaling and optimizing this for hybrid links, larger networks, and centralized vs.~decentralized control paradigms. 
Collectively, these efforts build a scalable, realistic foundation for deploying and evaluating next-generation quantum networks.

Future work includes integrating devices and error correction models into SeQUeNCe and comparing centralized vs.~distributed control schemes in the control and service plane of quantum networks.
We will conduct studies on hybrid quantum error management strategies similar to~\cite{pathumsoot2024hybridqem}. While the authors focused on linear repeater chains of trapped ions with eight hops maximum, we will scale our simulations to complex topologies considering hundreds of heterogeneous nodes and the protocols developed in \Cref{sec:qec}.
As described in~\cite{beauchamp2025whitepaperquantuminternet}, we are also interested in investigating under what conditions services level agreements  between users and the network are required and what behaviors can be observed under different traffic patterns and network topologies.

\section{Integration} \label{sec:integration}
Integration efforts highlight the co-design aspects of the InterQnet project (see \Cref{fig:concept}).
Our current work focuses on the coexistence of quantum and classical signals, stabilization of quantum channels under realistic fiber conditions, and orchestration of heterogeneous devices. 
These activities are essential to InterQnet-Achieve, where a three-node heterogeneous network will demonstrate entanglement distribution across superconducting, erbium, and ytterbium platforms, to address challenges \textbf{C1} and \textbf{C2}.
Moreover, results of these integration activities inform InterQnet-Scale by identifying key constraints and performance trade-offs for realistic architectures (see challenge \textbf{C3}), improving the realism of our models and increasing confidence in our simulations.

\begin{figure}
    \centering
    \includegraphics[width=\columnwidth]{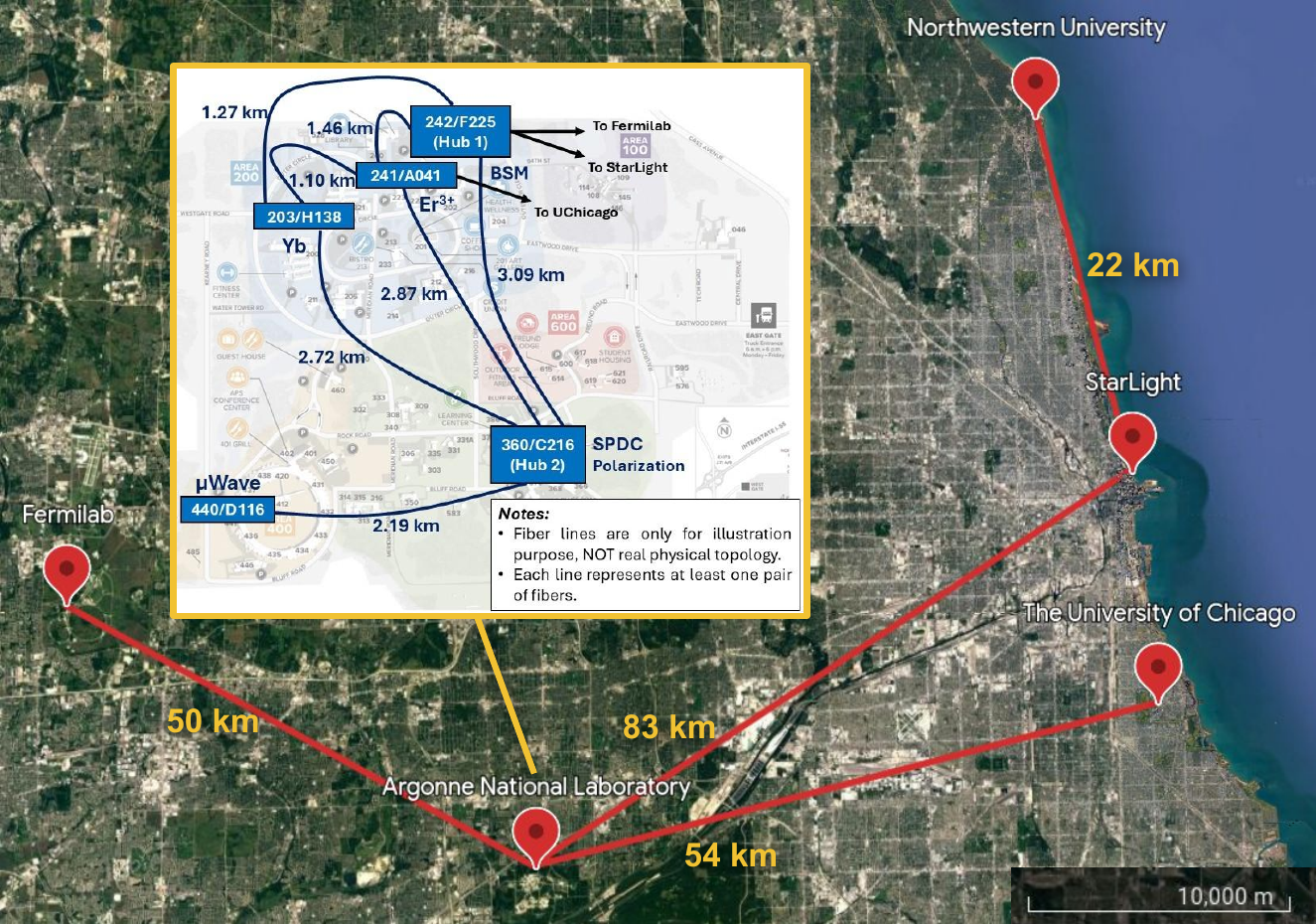}
    \caption{Dark fiber optics network in the great Chicago area, including Argonne and partner institutions. The inset shows Argonne's quantum network testbed (ARQNET), where InterQnet integration experiments are conducted.}
    \label{fig:arqnet-topo}
\end{figure}

\subsection{Quantum-Classical Coexistence over Deployed Fiber} \label{sec:coexistence}

We prototyped an on-demand entanglement distribution service over ARQNET (see inset in \Cref{fig:arqnet-topo}), addressing the challenge of spontaneous Raman scattering (SpRS) induced by coexisting classical signals~\cite{islam2024experiences}. We developed an empirical model that quantitatively describes the relationship between the coincidences-to-accidentals ratio (CAR) and pump power of entangled photon sources in the presence of classical traffic:

\begin{equation}
CAR = \frac{\alpha_s \cdot \alpha_i \cdot \mu_c}{((\mu_c + \mu_{sn}) \cdot \alpha_s + d_s)((\mu_c + \mu_{in}) \cdot \alpha_i + d_i)} + 1,
\end{equation}
where $\alpha_x$ denotes transmittance, $\mu_c$ represents correlated photon pairs per pulse, and $\mu_{xn}$ represents noise photons from SpRS. This framework enables operators (or a future network orchestrator) to optimize quantum device settings dynamically based on classical traffic levels and achieve a maximum CAR under the defined constraints.

\begin{figure}[htbp]
    \centering
\includegraphics[width=0.85\linewidth]{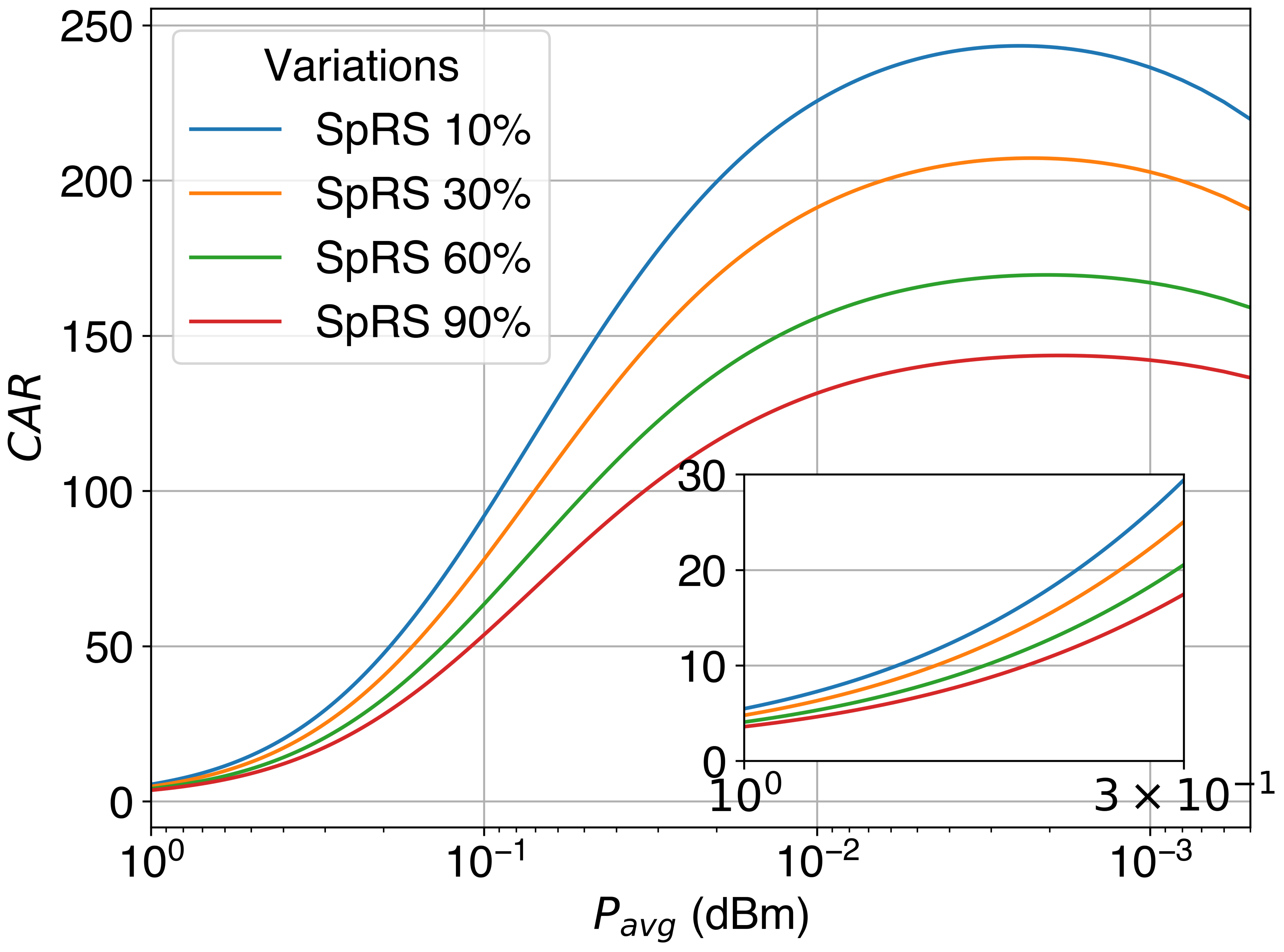}
    \caption{Dependence of the coincidences-to-accidentals ratio (CAR) on average power of the pump pulses for varying amounts of SpRS. Plot reproduced from~\cite{islam2024experiences}.}
    \label{fig:car_model}
\end{figure}

We established a proportional relationship between our theoretical model and the actual input power of an entangled photon source, which we subsequently utilized to execute experimental demonstrations of the on-demand entanglement service. Experimental validation was conducted on ARQNET utilizing deployed fiber connections between buildings 360 and 440, covering 4.38 kilometers (round trip). The experimental setup integrates classical traffic generation using 10GBASE-LR transceivers at 1330~nm with a commercial entangled photon source (NuCrypt EPS-1000-W) operating at 1550.10~nm. Classical and quantum signals are multiplexed using wavelength division multiplexers and routed through a software-controlled optical cross-connect, enabling both co-propagation and counter-propagation scenarios (only co-propagation is presented).

\begin{figure}[htbp]
    \centering
    \includegraphics[width=0.99\linewidth]{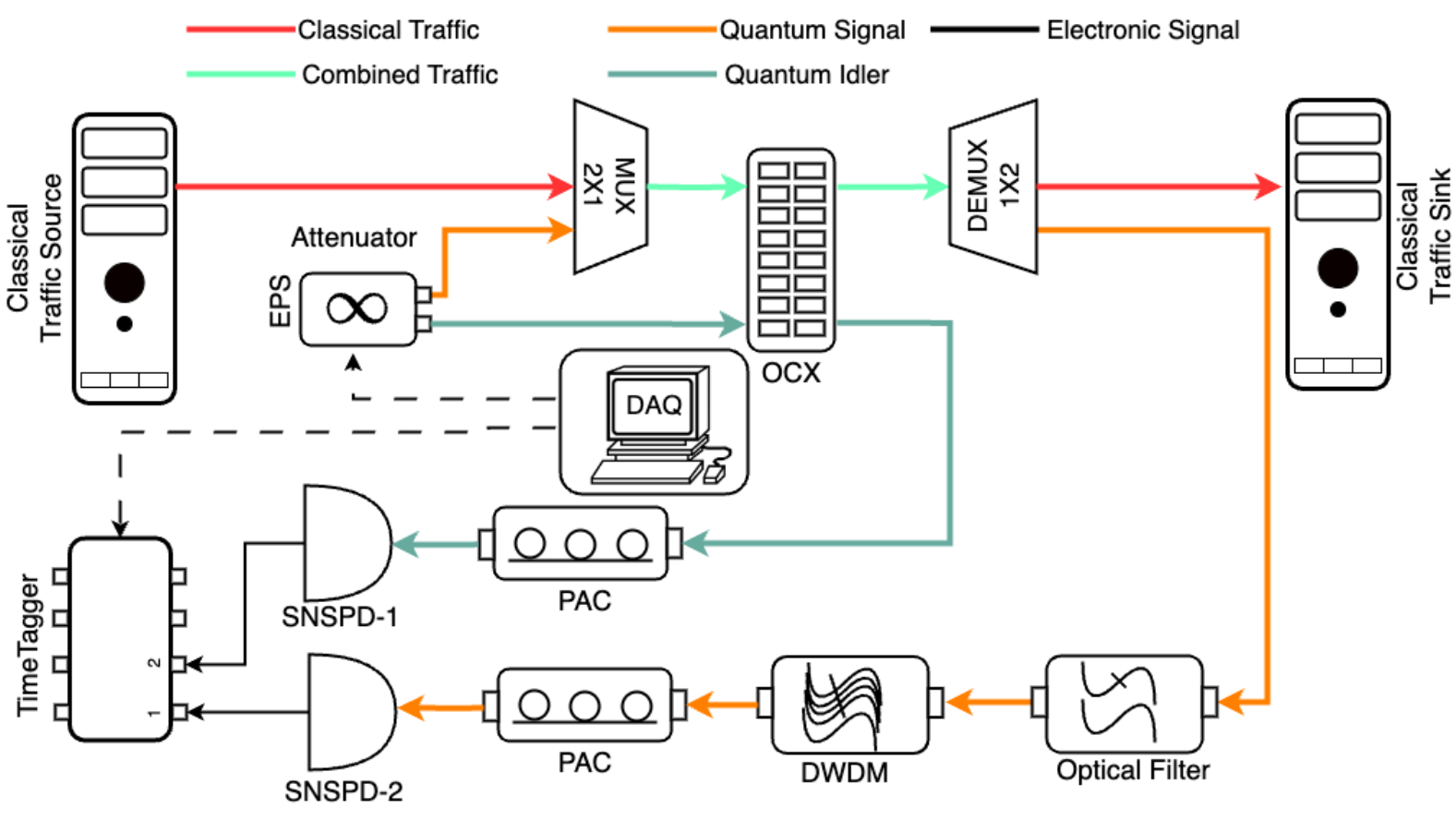}
    \caption{Schematic of the experimental setup in~\cite{islam2024experiences}, showing the integration of classical and quantum traffic over deployed fiber infrastructure with multiplexing and filtering components.}
    \label{fig:experimental_setup}
\end{figure}

Entanglement quality was validated through two-photon interference measurements using a 1~ns coincidence window, achieving coincidence rates of 155--120 counts per second. Fringe visibilities calculated as $V = (N_{max} - N_{min})/(N_{max} + N_{min})$ ranged up to 68\% and remained within a similar range with or without classical light being present in the system. Our simulations and preliminary tests at ARQNET demonstrate the model's capability to optimize quantum network performance by guiding configuration parameters such as EPS pump power, detector bias, and classical signal intensity. However, increasing the number of automated variables may introduce additional forecasting complexities for the model.

\begin{figure}[htbp]
    \centering \includegraphics[width=0.99\linewidth]{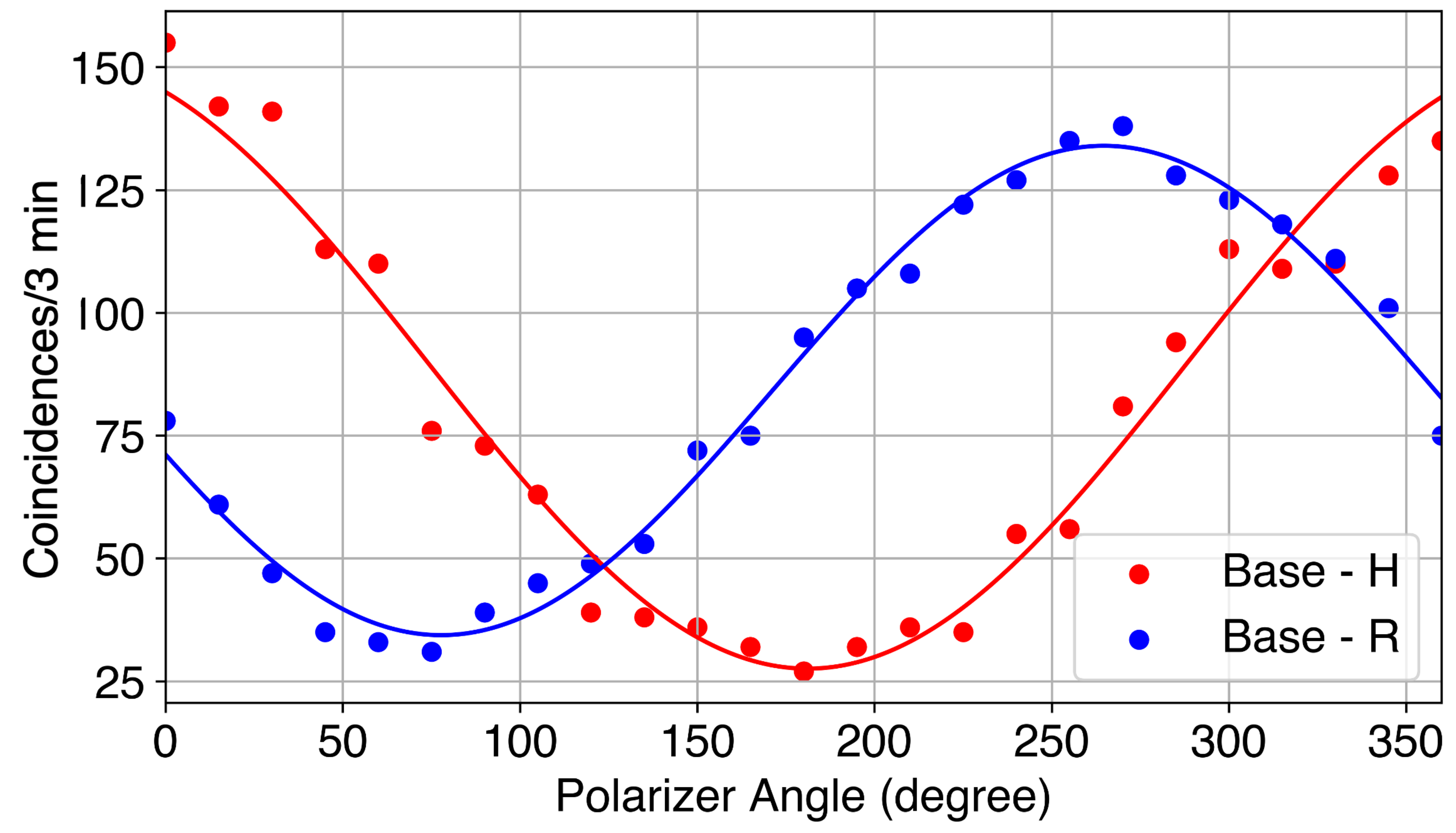}
    \caption{Two-photon interference fringes showing maintained entanglement quality with visibilities of 59--68\% in the presence of coexisting classical traffic. Reproduced from~\cite{islam2024experiences}.}
    \label{fig:interference_fringes}
\end{figure}

These results validate the feasibility of automated entanglement distribution services where users can request specific CAR values and the system dynamically adjusts pump power settings based on the predictive model. The successful coexistence of quantum and classical signals over deployed fiber infrastructure establishes a foundation for scalable quantum networking architectures integrated with existing telecommunications infrastructure.
We also demonstrated our quantum network's capability to achieve Hong--Ou--Mandel interference between photons generated at two distinct sources and transmitted through deployed fiber, alongside a coexisting clock signal generated by transceivers running at 2~Gbps \cite{RAMESH2025}. 
We verified this interference between weak coherent-state photons and photons from a heralded single-photon source located remotely.
A maximum dip visibility of 0.58 $\pm$ 0.04 was achieved. 
The sources were connected via 4.3~km of deployed fiber and had a coexisting clock signal in the same fiber to synchronize them. This represents a first step toward achieving teleportation over our network and demonstrates that it maintains sufficient timing and polarization stability required for scalable quantum networking.

\subsection{Polarization Stabilization Using Machine Learning} \label{sec:pol-ml}
Random variations in the birefringent profile of optical fibers---due to both static manufacturing defects and dynamic environmental processes---cause rotations to the state of polarization (SOP) of the electric field. While orthogonal states will remain orthogonal, in general, the polarization state at the output end of a fiber channel will not match the one sent at the input channel. Thus, in order to properly interfere quantum information photons encoded in polarization, two nodes must align their polarization bases so that they match.

\begin{figure}[thbp]
    \centering
    \includegraphics[width=0.99\linewidth]{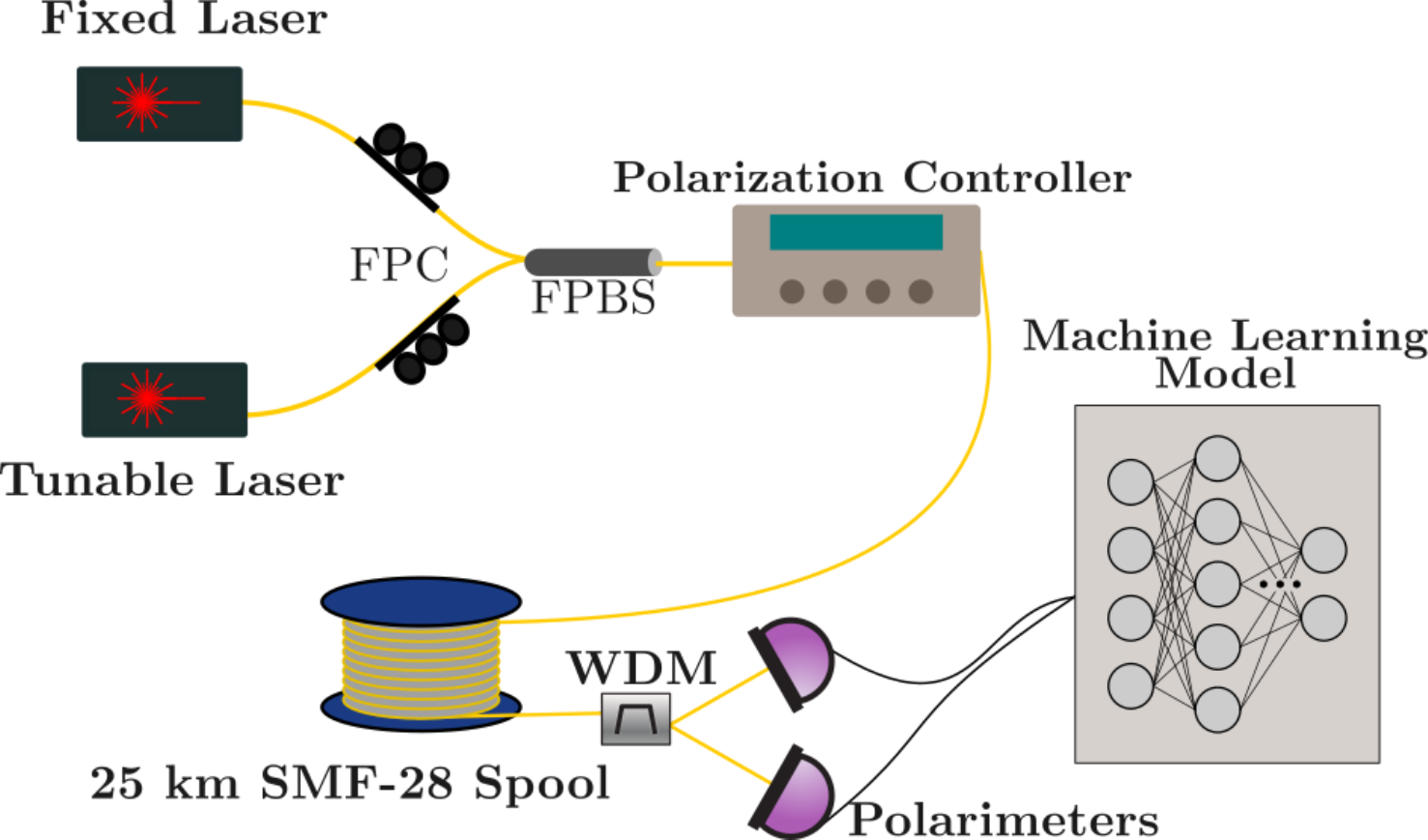}
    \caption{Experimental setup used in~\cite{Eastman25} to collect SOP polarization data for machine learning. The polarization controller acts as a scrambler, and the fiber spool is used to add the effect of polarization mode dispersion.}
    \label{fig:pol_ml_setup}
\end{figure}

\begin{figure}[thbp]
    \centering
    \includegraphics[width=\linewidth]{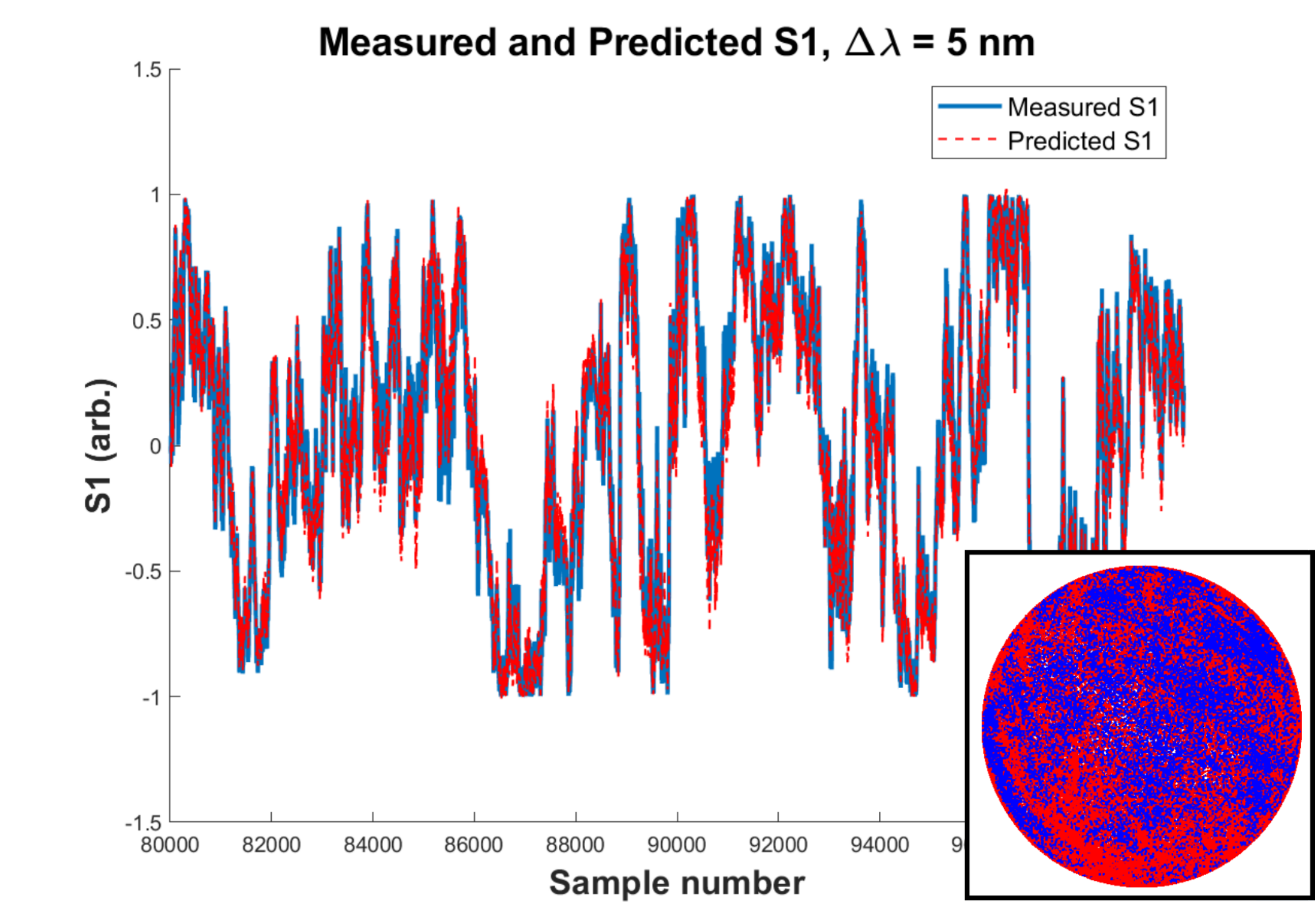}
    \caption{Main: Measured S1 parameter and our model predictions. The model faithfully tracks the state of polarization with a root mean square error of 0.072. Inset: Measured SOP states on the Poincar\'e sphere, demonstrating that we sample all possible SOPs. Reproduced from~\cite{Eastman25}.}
    \label{fig:pol_ml_result}
\end{figure}

We are currently studying the use of machine learning to find correlations between quantum information signals and classical control signals separated in wavelength. In the past, limitations caused by polarization mode dispersion (PMD) and the corresponding differential group delay ($\tau$), have required that control signals be placed within $\approx 2$~nm of their quantum counterparts \cite{xavier_full_2008}. We used an attention-based machine learning model to look for higher-order correlations between distantly separated signals. We generate training data for the model with a polarization scrambler, consisting of a motorized polarization controller to vary the SOP of the two signals followed by a 25~km spool of SMF-28 fiber to consider the effects of PMD (see \Cref{fig:pol_ml_setup}). After demultiplexing the two signals using a wavelength division multiplexer (WDM), we send each signal to a Stokes polarimeter, which reports each SOP as three Stokes parameters, which can be considered Cartesian coordinates on the Poincar\'e sphere. Our model directly predicts the SOP of the data signal based on the SOP of the control signal. Our most recent results~\cite{Eastman25} were able to predict the SOP of a signal separated from the control by 5~nm with an average root mean square error of 0.072 in individual Stokes parameters, as demonstrated in \Cref{fig:pol_ml_result}.

More flexibility is granted for allocating polarization control signals by allowing greater separation between the control and quantum signals. In a perfect scenario, any free WDM channel in the network can be used as a polarization control channel, so long as the machine learning model can accommodate the degree of separation between the two signals. Moreover, if a control channel needs to be reallocated, it can be moved to any other open channel without degrading the quality of polarization tracking. Furthermore, by training the model on a number of different wavelength separations, one might be able to stabilize multiple quantum channels with a single control signal, reducing the total number of control channels required in a network.

\subsection{Quantum Network Orchestrator Prototype}
\begin{figure}
    \centering
    \includegraphics[width=\columnwidth]{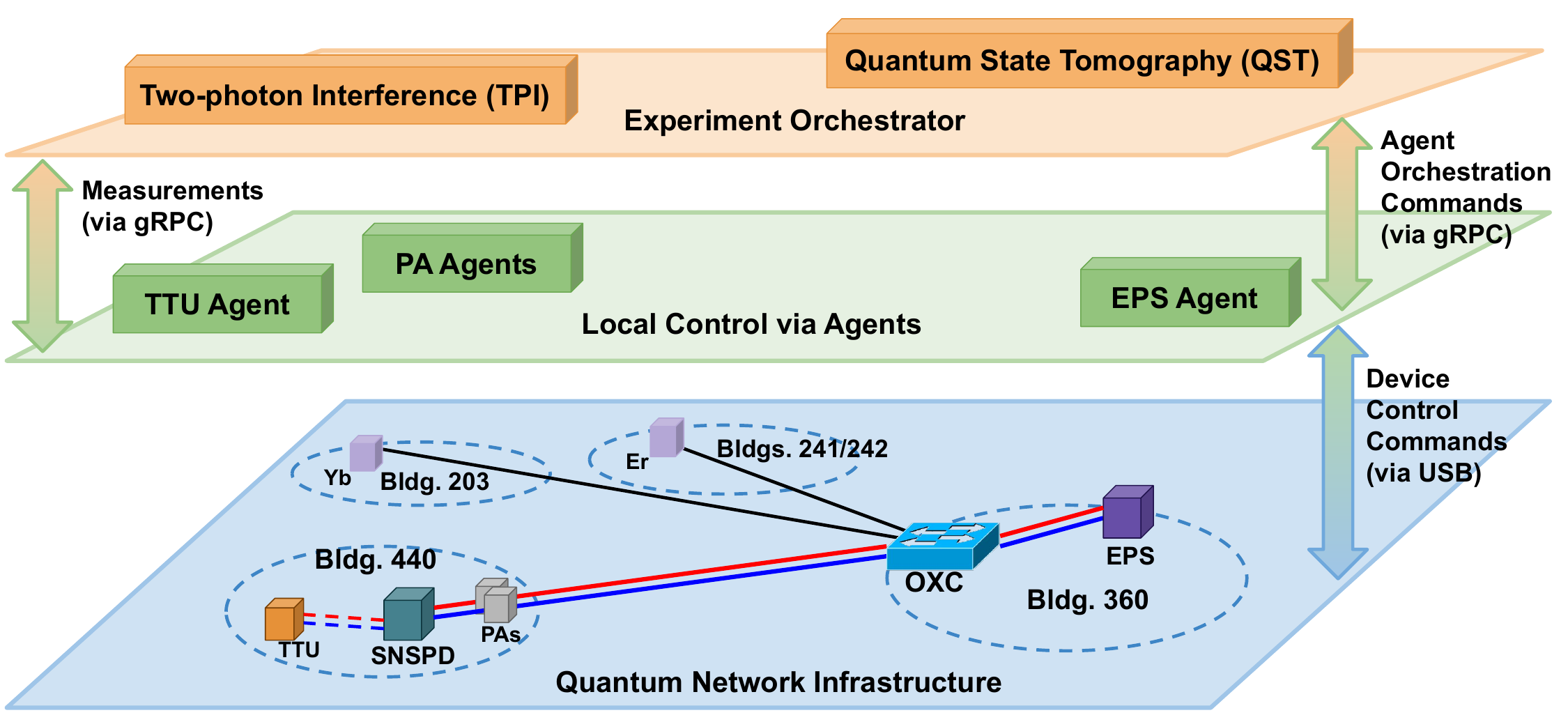}
    \caption{InterQnet orchestrator prototype following the plane abstraction.}
    \label{fig:orch}
\end{figure}

To coordinate heterogeneous quantum devices and protocols on ARQNET, we developed a modular orchestrator that instantiates the three-plane abstraction introduced in \Cref{sec:net-arch}. The orchestrator adopts a microservice architecture where each physical device (such as entangled photon sources, polarization analyzers, or time taggers) is encapsulated as a gRPC-based software agent. This design enables the orchestrator to issue high-level commands while device-specific control logic is handled locally, allowing protocols such as two-photon interference or quantum state tomography to be executed automatically across distributed nodes. \Cref{fig:orch} illustrates the orchestrator architecture, showing the gRPC-based interaction between device agents and the central control service.

A central challenge in distributed photonic networks is synchronization. Our prototype integrates a lightweight hybrid clock distribution scheme that simultaneously transmits 10~MHz and 1~PPS signals via radio-over-fiber, combined with NTP-based alignment. This approach provides picosecond-level timing coordination between remote Swabian time taggers without the complexity of GPS or White Rabbit systems, ensuring that coincidence detection and entanglement verification can be performed accurately across kilometers of fiber.

The orchestrator also incorporates polarization drift compensation as a gRPC service, enabling autonomous calibration and stabilization routines that replace manual realignment. 
Although our orchestrator follows the three-plane abstraction, the implementation is rooted in centralized control and SDN design principles. This approach aligns with current efforts from QUANT-NET~\cite{quantunet-qce25} and the Quantum Internet Alliance~\cite{beauchamp2025modularquantumnetworkarchitecture}.

\subsection{Summary of Accomplishments and Open Challenges}
Current integration efforts on ARQNET lay the foundation for the final InterQnet-Achieve demonstration.
The ARQNET topology imposes a real constraint on our design because fiber links connecting the laboratories hosting heterogeneous devices and the integrated BSM and QFC have only two fiber strands; thus the coexistence of quantum and classical traffic will be unavoidable.
Coexistence studies showed that entanglement can be maintained alongside classical traffic, supported by an empirical model of Raman scattering and validated experimentally over deployed fiber. 
Polarization stabilization experiments using machine learning demonstrated that quantum channels can be predicted from classical control channels, with the potential to reduce the number of wavelengths needed for robust operation under dynamic conditions. 
While the final heterogeneous devices in our testbed will rely on time-bin encoded photons, having polarization stabilization can improve entanglement rates.
A prototype quantum network orchestrator was also developed, implementing centralized control with serial drivers for device control and gRPC-based coordination.

Challenges remain in extending coexistence studies to more traffic patterns and longer spans and demonstrating polarization stability over extended timescales.
Data collected in ARQNET experiments can be used to validate SeQUeNCe models, as we have already shown in~\cite{Singal:24}.
Scaling the orchestrator to manage heterogeneous entanglement across superconducting, erbium, and ytterbium platforms will require continuous development and integration of the software framework. 
Our current prototype provided valuable information on the engineering challenges of deploying a centralized controller and agents distributed across an institutional LAN composed of multiple security domains (i.e., firewalls). More work is required on improving our abstractions toward integration of the orchestrator with the local control systems of our heterogeneous qubit platforms.

These integration efforts complement the error management strategies of \Cref{sec:qec} and the architectural analysis of \Cref{sec:net-arch} by providing real data to inform the models, advancing InterQnet toward its goals of heterogeneous demonstrations (Achieve) and guiding scalable design principles (Scale). The minimal configuration for InterQnet-Achieve considers a single-repeater link, with the Yb atom platform acting as the repeater and \Erppp\ and superconducting devices acting as end nodes (see path $\mu$Wave--Yb--Er in \Cref{fig:triangle-qnet}). Each elementary link will have a mid-station containing QFC devices and the BSM. Entanglement swapping at the Yb repeater will distribute entanglement between the end nodes. Future extensions of InterQnet-Achieve will pursue demonstrations of the entanglement distillation protocol presented in~\Cref{sec:qec-dual-species}.

\section{Conclusion} \label{sec:conclusion}
The path from small-scale demonstrations to national-scale, heterogeneous quantum networks remains blocked by challenges of scalability, control, and architectural clarity. Most existing efforts focus on homogeneous systems or simulations disconnected from experimental feedback. InterQnet distinguishes itself by: (i) integrating heterogeneous platforms, including atoms, ions, and superconducting qubits, in the first demonstrations of their kind; (ii) performing architectural simulations in SeQUeNCe that are parameterized by realistic device and network constraints, informed directly by experiments; (iii) pursuing full-stack co-design, where protocols, control logic, and device models are developed in tandem to enable end-to-end system exploration; and (iv) evaluating architectures under realistic conditions, exploring trade-offs in service time, fidelity, and resilience within fiber-deployed topologies rather than abstract models.

Through this combination of experimental integration and systems-level study, InterQnet has advanced toward building a full-stack heterogeneous quantum network (InterQnet-Achieve) and guiding the design of scalable architectures (InterQnet-Scale). This work establishes a foundation for future demonstrations such as teleportation and Bell inequality violations on ARQNET, as well as scaling qLDPC-based purification to larger networks. Together, these results move the field closer to the realization of robust, heterogeneous quantum networking at a national scale.

\bibliographystyle{IEEEtran}
\bibliography{bibliography}


\end{document}